\documentclass[twocolumn]{aastex631}

\usepackage{color}
\usepackage{amsmath}

\newcommand{\rev}[1]{{#1}}

\newcommand{\Msun}{\ensuremath{\mathrm{M}_\odot}}
\newcommand{\Rsun}{\ensuremath{\mathrm{R}_\odot}}

\shorttitle{Binary Precursors}
\shortauthors{Tsuna et al.}

\begin{document}

\title{Bright Supernova Precursors by Outbursts from Massive Stars with Compact Object Companions}

\correspondingauthor{Daichi Tsuna}
\email{tsuna@caltech.edu}

\author[0000-0002-6347-3089]{Daichi Tsuna}
\affiliation{TAPIR, Mailcode 350-17, California Institute of Technology, Pasadena, CA 91125, USA}
\affiliation{Research Center for the Early Universe (RESCEU), School of Science, The University of Tokyo, 7-3-1 Hongo, Bunkyo-ku, Tokyo 113-0033, Japan}

\author[0000-0002-9350-6793]{Tatsuya Matsumoto}
\affiliation{Department of Astronomy, Kyoto University, Kitashirakawa-Oiwake-cho, Sakyo-ku, Kyoto, 606-8502, Japan}
\affiliation{Hakubi Center, Kyoto University, Yoshida-honmachi, Sakyo-ku, Kyoto, 606-8501, Japan}
\affiliation{Department of Physics and Columbia Astrophysics Laboratory, Columbia University, Pupin Hall, New York, NY 10027, USA}

\author[0000-0003-2872-5153]{Samantha Chloe Wu}
\affiliation{TAPIR, Mailcode 350-17, California Institute of Technology, Pasadena, CA 91125, USA}

\author[0000-0002-4544-0750]{Jim Fuller}
\affiliation{TAPIR, Mailcode 350-17, California Institute of Technology, Pasadena, CA 91125, USA}
 
\begin{abstract}
A fraction of core-collapse supernovae (SNe) with signs of interaction with a dense circumstellar matter are preceded by bright precursor emission. While the precursors are likely caused by a mass ejection before core-collapse, their mechanism to power energetic bursts, sometimes reaching $10^{48}$--$10^{49}\ {\rm erg}$ that are larger than the binding energies of red supergiant envelopes, is still under debate. Remarkably, such a huge energy-deposition should result in an almost complete envelope ejection and hence a strong sign of interaction, but the observed SNe with precursors show in fact typical properties among the interacting SNe. More generally, the observed luminosity of $10^{40-42}\,\rm erg\,s^{-1}$ is shown to be challenging for a single SN progenitor. To resolve these tensions, we propose a scenario where the progenitor is in a binary system with a compact object (CO), and an outburst from the star leads to a super-Eddington accretion onto the CO. We show that for sufficiently short separations, outbursts with moderate initial kinetic energies of $10^{46}$--$10^{47}$ erg can be energized by the accreting CO so that their radiative output can be consistent with the observed precursors. We discuss the implications of our model in relation to CO binaries detectable with \textit{Gaia} and gravitational wave detectors.
\end{abstract}


\section{Introduction}
Mass loss is an important process in massive stars that characterizes their evolution and the appearances of their final supernova (SN) explosions \citep[e.g.,][]{Smith2014}. A wide variety of core-collapse explosions show signatures of dense circumstellar matter (CSM) around the progenitor, indicating suddenly enhanced mass-loss months to centuries prior to core-collapse. These span from SNe with narrow lines classified as Type IIn/Ibn/Icn \citep{Schlegel1990,Pastorello08,Gal-Yam22}, and possibly Type II-P/L SNe that comprise about half of core-collapse SNe \citep{Moriya11,Khazov16,Morozova17,Morozova18,Yaron17,Forster18,Bruch21,Bruch23}.

For a fraction of these SNe, the dense CSM is linked to bright optical flares observed from months to years before core-collapse, often called SN precursors. Since the first finding of a precursor in a Type Ibn SN 2006jc two years before the SN \citep{Nakano_et_al_2006,Pastorello07}, precursors have been observed in many Type IIn SNe \citep[e.g.,][]{Mauerhan+2013,Fraser+2013b,Ofek+2013b,Ofek+2014c,Margutti+2014,EliasRosa+2016,Strotjohann21,Fransson22,Hiramatsu+2023} and a few SNe of other types including Type Ibn SN 2019uo \citep{Strotjohann21} and Type II-P SN 2020tlf \citep{Jacobson-Galan22}. \rev{A recent study with a large sample of interacting SNe \citep{Strotjohann21} finds that bright precursors months before core-collapse occur in $5$--$69~\%$ of Type IIn SNe and $\sim 10~\%$ of Type Ibn SNe.} Understanding these precursors is important as clues to probing the mass-loss mechanism that produces dense CSM in the final years of massive stars.

The observed precursors are bright with luminosities of $10^{40}$--$10^{42}$ erg s$^{-1}$, which are orders of magnitude higher than the Eddington limit of massive stars. While the super-Eddington energy injection may naturally explain the generation of dense CSM \citep[e.g.,][]{Matsumoto22}, the mechanism that triggers such powerful energy injection is yet unclear. For the wave-driven mass loss mechanism where specific predictions exist, the luminosity of the (possible) outbursts months to years before core-collapse are estimated to be only $\lesssim 10^6L_\odot$ (\citealt{Shiode&Quataert2014,Fuller2017,Fuller&Ro2018,Leung+2021b,Wu22a}, see also discussion in \citealt{Strotjohann21}). Model-agnostic radiation hydrodynamical simulations of partial envelope ejection triggered by energy injection from the core, aimed to phenomenologically reproduce the CSM of interacting SNe, find a similar upper limit in the precursor luminosity \citep[e.g.,][]{Quataert+2016,Kuriyama&Shigeyama2020,Kuriyama&Shigeyama2021,Tsang22,Tsuna23}.

\begin{figure*}
    \centering \includegraphics[width=\linewidth]{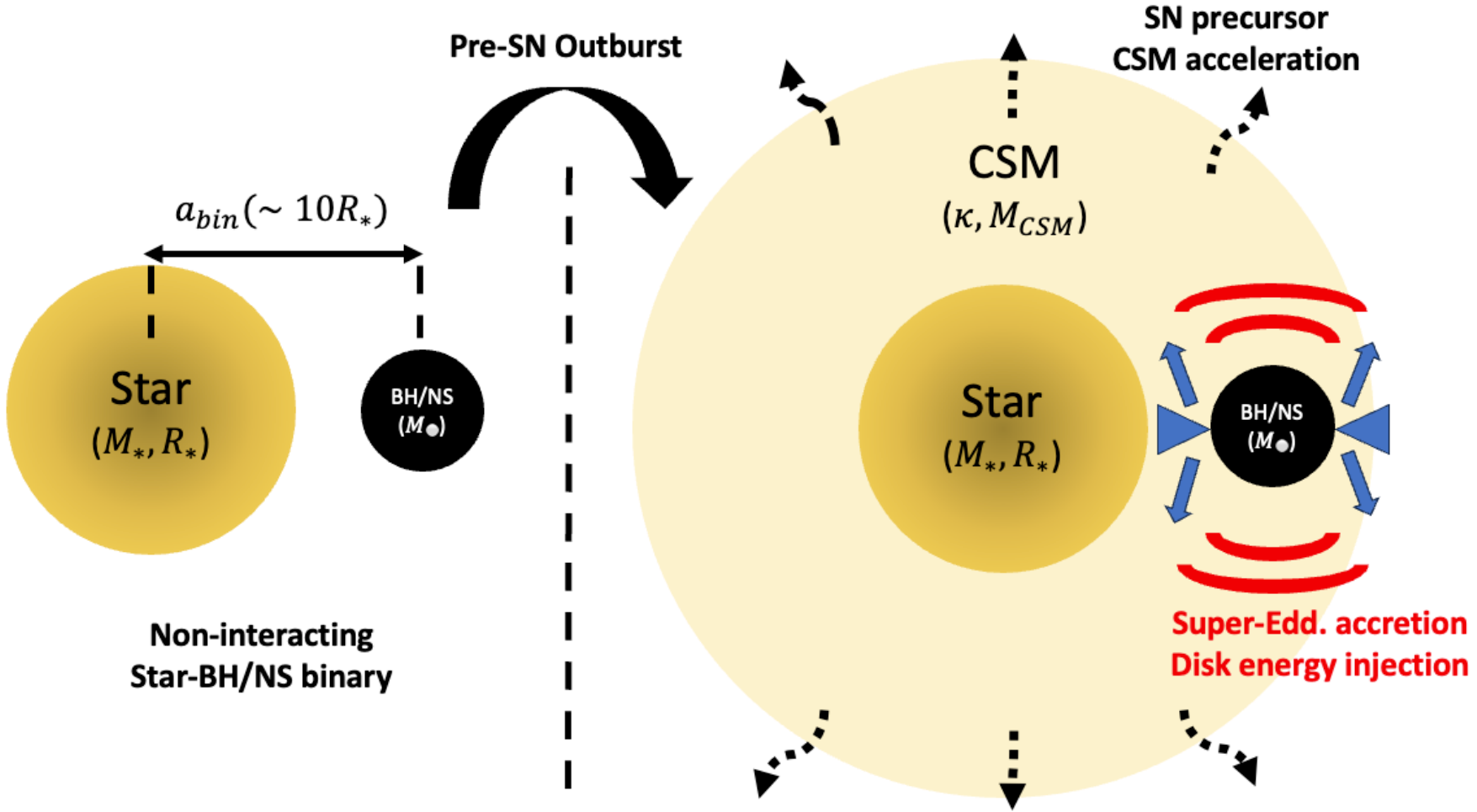}
    \caption{Schematic picture of our model (not to scale). The SN progenitor in a close binary system with a CO undergoes a mass outburst that creates the dense CSM. This generally results in super-critical accretion of a fraction of the CSM onto the CO, which leads to outflows that boost the energy of the CSM and power the precursor emission.}
    \label{fig:schematic}
\end{figure*}

In this work, we suggest that the observed bright precursors can be reproduced if the progenitor is in a binary system with a compact object (CO) companion, i.e., a stellar-mass black hole (BH) or a neutron star (NS). Most of young massive stars ($>70\%$) live as interacting binaries \citep{Sana12}, and a significant fraction of them can result in such star-CO systems. Binary interaction is also considered to be the main formation channel for stripped-envelope SNe of Type Ibc \citep[e.g.,][]{Shigeyama90,Podsiadlowski92,Eldridge08,Smith11,Chen23}, where the companions can be COs if they were the more massive of the initial binary and underwent core-collapse first.

The schematic picture of our model is shown in Figure \ref{fig:schematic}. The erupted material that encounters the CO's Bondi sphere would be accreted, forming a disk around it. For binaries close enough the accretion can become super-Eddington and launch an energetic outflow, which can lead to additional energy injection into the remaining CSM \citep[e.g.,][]{Dexter13,Kimura17,Moriya18_fallback}. This can boost the energy budget of the precursor \citep[e.g.,][]{Mcley14,Danieli19}, drastically reducing the required energy deposition for the partial eruption to values in line with proposed theoretical models.

In Section \ref{sec:single_lum} we first elaborate on the difficulty of reproducing the observed bright precursors by eruptions of single massive stars. Then in Section \ref{sec:model} we describe our detailed modeling of precursors from massive star-compact object (CO) binary systems, and in Section \ref{sec:result} demonstrate that such systems can generally explain the energetics of the observed precursor emission. In Section \ref{sec:discussion} we discuss the formation channel of these binaries, as well as ways to test the model with independent observations. We conclude in Section \ref{sec:conclusion}.

\section{Precursors from Single Stars}
\label{sec:single_lum}
In this section, we estimate the radiative output of a single mass eruption event from an isolated (non-binary) star, and show its difficulty to reproduce the observed precursor events. 

The radiated energy in the observed precursors, of $10^{47}$--$10^{49}$ erg and typically of the order of $10^{48}$ erg (see e.g., Table 4 of \citealt{Strotjohann21} and Table 1 of \citealt{Matsumoto22}), requires that at least a comparable energy is injected to the envelope. This disfavors an eruption from a red supergiant (RSG) progenitor, which would result in (near-complete) ejection of the hydrogen-rich envelope with its small binding energy of
\begin{eqnarray}
    E_{\rm bind} &\sim& \frac{GM_*M_{\rm env}}{R_*} \nonumber \\
                    & \sim & 4\times 10^{47}\ {\rm erg} \left(\frac{M_*}{10 \! \ M_\odot}\right) \! \left(\frac{M_{\rm env}}{5 \! \ M_\odot}\right) \! \left(\frac{R_*}{500 \! \ R_\odot}\right)^{\!-1}.
                    \label{eq:Ebind}
\end{eqnarray} 
Here $G$ is the gravitational constant, and $M_*$, $M_{\rm env}$ and $R_*$ are respectively the total mass, envelope mass, and radius of the star. Such a near-complete envelope ejection is unlikely for the observed precursors, because (i) the mass of the CSM constrained from the rise time of the subsequent SN is typically $\lesssim$ a few $M_\odot$ and (ii) Type IIn SNe with precursors are not on the luminous end of the entire IIn population \citep{Strotjohann21}. For the former, the rise times for these SN samples are estimated to be $t_{\rm rise}<25$ days \citep[][Table 4]{Strotjohann21}, and the timescale of radiative diffusion from the swept-up CSM being shorter than this limits the mass of the CSM as \citep[e.g.,][]{Moriya15}
\begin{eqnarray}
    M_{\rm CSM}&\lesssim&\frac{4\pi ct_{\rm rise}(v_{\rm sh}t_{\rm rise})}{3\kappa} \nonumber \\
    &\approx& 3\ M_\odot\left(\frac{t_{\rm rise}}{25\ {\rm day}}\right)^2 \left(\frac{\kappa}{0.3\ {\rm cm^2\ g^{-1}}}\right)^{-1} \ ,
\end{eqnarray}
where we assume a rather large SN shock velocity of $v_{\rm sh}=10^4$ km s$^{-1}$, $c$ is the speed of light, and $\kappa$ is the opacity assumed to be constant in time and radius. For the latter, the peak magnitude of SNe with precursors span from $-19$ to $-17$ mag \citep[roughly correspoding to $3\times10^{42}-10^{43}\,\rm erg\,s^{-1}$,][]{Strotjohann21,Matsumoto22}, which agrees with the range seen in typical Type IIn SNe \citep{Kiewe+2012,Ofek14_rise_time_peak,Nyholm+2020}. This indicates that the mass of CSM for SNe with bright precursors should be comparable to those without, and should not be biased towards higher masses. 

For Type Ibn/Icn SNe from helium (He)/Wolf-Rayet (WR) stars, light curve modelling of these SNe in general suggests a CSM mass of $\sim 0.01$--$1\ M_\odot$ \citep{Maeda22,Dessart+2022,Wu22b,Takei24}. \rev{For these stripped progenitors we expect a broad range of masses in their outer helium layers ($\sim 0.1$-- 2$M_\odot$ for solar metallicity; e.g., \citealt{Yoon_Woosley_Langer10,Yoon17}). Thus for SNe Icn (and some SNe Ibn) complete envelope ejection can still be compatible with observations, though energetically more demanding than partial ejection.}

This energetics problem, on the difficulty to produce the observed precursor's energy without significantly modifying the progenitor's structure, can be avoided if the progenitor is more compact than RSGs, such as blue supergiants (luminous blue variables) or He/WR stars. However, even in these cases, a single episode of partial mass ejection struggles to reproduce the bright luminosity of the precursors because of losses to adiabatic expansion, as we show below.

Suppose that the precursor event ejects a CSM of mass $M_{\rm CSM}$ with a velocity 
$v_{\rm CSM}=\eta v_{\rm esc}$ (here $\eta$ is a parameter) and
\begin{eqnarray}
    v_{\rm esc} = \sqrt{\frac{2GM_*}{R_*}} \approx 90\ {\rm km\ s^{-1}}  \left(\frac{M_*}{10\ M_\odot}\right)^{1/2}\left(\frac{R_*}{500\ R_\odot}\right)^{-1/2}
\end{eqnarray}
is the surface escape velocity. Under the assumption that the energy is sourced from near or inside the base of the envelope\footnote{There are scenarios for injecting energy close to the surface, such as wave heating in the case of stripped stars \citep[e.g.,][]{Fuller&Ro2018,Leung+2021b}. However, in this scenario energy injection is continuous (i.e. longer than the dynamical time near surface), and such steady-state energy injection driving a wind-like CSM have difficulties to power emission highly exceeding the Eddington luminosity \citep{Quataert+2016,Matsumoto22} unless strong velocity gradients develop internal shocks in the wind acceleration region.
}, for a partial ejection of the envelope we expect that the bulk CSM velocity is comparable to the surface escape velocity, i.e. $\eta\sim 1$ \citep{Linial+2021}.

Similar to SNe \citep{Arnett1980}, as the CSM expands and its optical depth drops, photons in the CSM escape to power the precursor. Its timescale is governed by the diffusion through the expanding CSM, which is
\begin{eqnarray}
    t_{\rm prec,0} &\approx& \sqrt{\frac{3\kappa M_{\rm CSM}}{4\pi c v_{\rm CSM}}} \nonumber \\
    &\approx& 80\ {\rm day} \left(\frac{\kappa}{0.3\ {\rm cm^2\ g^{-1}}}\right)^{1/2} \nonumber \\
    &&\times \left(\frac{M_{\rm CSM}}{0.1\ M_\odot}\right)^{1/2}\left(\frac{v_{\rm CSM}}{100\ {\rm km\ s^{-1}}}\right)^{-1/2}.
    \label{eq:t_prec0}
\end{eqnarray}
The internal and kinetic energy $E_{\rm kin}=M_{\rm CSM}v_{\rm CSM}^2/2$ are comparable upon the eruption, but until $t=t_{\rm prec,0}$ the internal energy is converted to kinetic energy by adiabatic expansion and reduced by a factor\footnote{This factor is under the assumption of radiation-dominated gas. In case gas pressure is important the energy budget of radiation decreases, only strengthening the conclusion of this section.} $\epsilon_{\rm rad}\approx (R_*/v_{\rm CSM}t_{\rm prec,0})\leq 1$. The luminosity of the precusor is then
\begin{eqnarray}
    L_{\rm prec,0}\approx \frac{\epsilon_{\rm rad} E_{\rm kin}}{t_{\rm prec,0}} \approx \eta^2 \frac{4\pi GM_*c}{3\kappa} = \frac{\eta^2}{3} L_{\rm Edd}
    \label{eq:Lprec_single}
\end{eqnarray}
where 
\begin{equation}
L_{\rm Edd}=1.7\times 10^{39}\ {\rm erg\ s^{-1}} \left(\frac{M_*}{10\ M_\odot}\right)\left(\frac{\kappa}{0.3\ {\rm cm^2\ g^{-1}}}\right)^{-1}
\end{equation}
is the Eddington luminosity.

For a typical $10\ M_\odot$ star, the fact that $\eta$ cannot be much greater than unity limits the luminosity of the eruption to the order of $10^{39}\ {\rm erg\ s^{-1}}$, which is much dimmer than the observed precursors with luminosities of $10^{40}$--$10^{42}$ erg s$^{-1}$. 
This scaling in the luminosity is also roughly seen in more detailed light curve modelling. For instance, radiation hydrodynamical simulations of partial envelope ejections find a luminosity of $10^{38}$--$10^{39}$ erg s$^{-1}$ for the resulting transient \citep{Fuller2017,Kuriyama&Shigeyama2020,Tsuna23}. A semi-analytical model by \cite{Matsumoto22} finds that for a RSG of $M_*=10M_\odot,\, R_*=10^3R_\odot$, a precursor luminosity of $L_{\rm prec}\approx 10^{41}$ erg s$^{-1}$ requires an eruption with CSM velocity of $\approx 400$--$700$ km s$^{-1}$ for $M_{\rm CSM}=0.1$--$10\ M_\odot$ (their Figure 4), which corresponds to $\eta\sim 10$.

We can consider two ways to break the limit on the luminosity (equation~\ref{eq:Lprec_single}), which is similar to the models proposed for super-luminous SNe \citep[e.g.,][]{Moriya18_SLSN,Gal-Yam19,Nicholl21}. The first possibility is to invoke a pre-existing CSM surrounding the progenitor at the time of the eruption. The collision of the two CSM shells can (re-)convert the kinetic energy into internal energy, effectively raising $\epsilon_{\rm rad}$. This however still requires an energy budget for $E_{\rm kin}$ of at least the radiated energy in the precursors, of $10^{47}$--$10^{49}$ erg \citep{Strotjohann21,Jacobson-Galan22}, which again is disfavored for RSG eruptions. Even for more compact progenitors, the mechanism that can realize this large energy deposition is unclear. Recent models of wave heating find that energy injections of $\sim \! 10^{48}$ ergs are realized in a limited set of progenitors only during the end stages of nuclear burning ($\lesssim 0.1$ yr before core-collapse; \citealt{Wu&Fuller2021,Leung+2021b}), while the precursors are typically seen several months to years before the SN.

The second possibility we explore in the following is a binary with a CO, which acts as a ``central engine" to power the light curve. For binaries with sufficiently close separations, the gravitational trapping of a fraction of the CSM by the CO can lead to super-Eddington accretion. As we show below, this accretion can generate outflows that carry energy greatly exceeding the original outburst energy, enhancing the energy budget in equation \ref{eq:Lprec_single}. Thus this scenario is an energetically favorable solution compared to the scenario invoking CSM interaction.

\section{Precursors from Binaries with Compact Object Companions}
\label{sec:model}
We consider a circular binary system with separation $a_{\rm bin}$, composed of a CO of mass $M_{\bullet}$ and a massive star of mass $M_*$ that undergoes a partial envelope ejection. We restrict the separation to larger than that for Roche-lobe overflow to occur, i.e. $a_{\rm bin}\geq 3R_*$ \citep{Eggleton83} for mass ratios of $M_*/M_\bullet\gtrsim 0.3$ of interest in this work, which results in non-interacting binaries until the pre-SN outburst \rev{(for discussion on interacting binaries see Section \ref{sec:future_work})}. 

We estimate the properties of the energy injection from the CO, and calculate the time-dependent emission from the CSM. The formulations of this model are similar to \cite{Kimura17}, but we newly construct a one-zone light curve model including time-dependent effects of the accretion and ionization of the CSM. The model parameters we employ are summarized in Table \ref{tab:params}. The details of each parameter will be described in the following sections.

\begin{table*}[]
    \centering
    \begin{tabular}{cccc}
         \hline
         Parameters & RSG-BH(NS) & HeHighMass-BH(NS) & HeLowMass-BH(NS) \\
         \hline
         CO mass ($M_{\bullet}$) & &  $10M_\odot$ ($1.4M_\odot$) & \\
         Star mass ($M_*$) & $10M_\odot$ & $5M_\odot$ & $3M_\odot$ \\
         Stellar radii ($R_*$) & $500R_\odot$ & $5R_\odot$ & $50R_\odot$ \\
         Opacity ($\kappa$) & $0.3$ cm$^2$ g$^{-1}$ & $0.1$ cm$^2$ g$^{-1}$ & $0.1$ cm$^2$ g$^{-1}$ \\
         Ionization temperature ($T_I$) & $6000$ K & $10000$ K & $10000$ K \\
         Initial CSM velocity in units of $v_{\rm esc}$ ($\xi$) & & 2 & \\
         CSM mass ($M_{\rm CSM}$) & 0.1--3$M_\odot$ &  0.01--1$M_\odot$ & 0.01--1$M_\odot$ \\
         Binary separation ($a_{\rm bin}$) & &  (3--30)$R_*$ & 
    \end{tabular}
    \caption{Parameters adopted in our precursor model. Values shown in only one column are common to all three systems.}
    \label{tab:params}
\end{table*}

\subsection{Accretion Disk Wind from the CO Companion}

We assume that the CSM is expanding homologously and has a uniform density profile for simplicity.
Hereafter, we define the velocity as that in the rest frame of the star. The CSM has a velocity ranging from $v_{\rm min}\equiv \chi v_{\rm esc}$ to $v_{\rm max}\equiv \xi v_{\rm esc}$, where the minimal velocity is set by the condition that the CSM reaches the CO without falling back to the progenitor:
\begin{eqnarray}
\chi=\sqrt{1-R_*/a_{\rm bin}}\ .  
    \label{eq:chi}
\end{eqnarray}
This definition of $\chi$ is approximate as we ignore the gravity from the CO, but for a uniform density profile varying $\chi$ would lead to only slight changes in the density for a fixed CSM mass. We treat $\xi\gtrsim1$ as a free parameter not much larger than unity, based on the arguments in Section \ref{sec:single_lum}.

Given the velocity range, the (uniform) CSM density at time $t$ from the eruption is given by 
\begin{eqnarray}
    \rho_{\rm CSM}&=&
    \frac{3M_{\rm CSM}}{4\pi (v_{\rm esc}t)^{3} (\xi^{3}-\chi^{3})} \\
    &\sim& \frac{7\times 10^{-11}}{\xi^3-\chi^3}\ {\rm g\ cm^{-3}\,}\left(\frac{M_{\rm CSM}}{0.1\ M_\odot}\right) \nonumber \\
    &&\times \left(\frac{v_{\rm esc}}{100\ {\rm km\ s^{-1}}}\right)^{-3} \left(\frac{t}{100\ {\rm day}}\right)^{-3}.
\end{eqnarray}

The CSM starts to be accreted when it reaches the CO at at time
\begin{eqnarray}
t_{\bullet}&\equiv&\frac{a_{\rm bin}}{v_{\rm max}} \nonumber \\
&\sim&\frac{200}{\xi}{\,\rm day\,}\left(\frac{a_{\rm bin}}{5R_*}\right)\left(\frac{R_*}{500\,\Rsun}\right)\left(\frac{v_{\rm esc}}{100\,\rm km\,s^{-1}}\right)^{-1}\ .
    \label{eq:t_dot}
\end{eqnarray}
The gravitational pull of the CO traps the CSM (which is moving approximately homologously at a velocity of $a_{\rm bin}/t$) within the CO's Bondi radius
\begin{eqnarray}
    r_{\rm B}=\frac{GM_{\bullet}}{V^2}
    \approx1\times 10^{13}\ {\rm cm}\left(\frac{M_{\bullet}}{10\ M_\odot}\right)\left(\frac{V}{100\,\rm km\,s^{-1}}\right)^{-2}
\end{eqnarray}
where we define 
\begin{eqnarray}
    V&\equiv&\sqrt{(a_{\rm bin}/t)^2+v_{\rm orb}^2}\ ,
        \label{eq:V}\\ 
    v_{\rm orb}&=&\sqrt{{G(M_*+M_{\bullet})}/{a_{\rm bin}}} \nonumber \\
    &\approx& 40\ {\rm km\ s^{-1}} \left(\frac{M_*+M_{\bullet}}{20\ M_\odot}\right)^{1/2} \nonumber \\
    &&\times \left(\frac{a_{\rm bin}}{5R_*}\right)^{-1/2}\left(\frac{R_*}{500\,\Rsun}\right)^{-1/2}\ .
    \label{eq:v_orbital}
\end{eqnarray}
Equation (\ref{eq:v_orbital}) is the orbital velocity of the CO with respect to the star.
Note that when the CSM reaches the CO at $a_{\rm bin}$, the internal energy in the CSM is much smaller than the kinetic energy, and hence the contribution from the sound speed in the CSM to $V$ can be neglected. 

As the homologously expanding CSM contains velocity asymmetry of $\sim \! r_{\rm B}/t$ due to a velocity difference over the Bondi radius, it carries a specific angular momentum of $j \sim r_{\rm B}(r_{\rm B}/t)$. This results in a formation of an accretion disk around the CO with outer radius \citep{Shapiro76}
\begin{eqnarray}
    r_{\rm disk} &\approx& \frac{j^2}{GM_{\bullet}} \nonumber \\
    &\sim& 8\times 10^{10}\xi^2{\,\rm cm\,}\left(\frac{M_\bullet}{10\,\Msun}\right)^3\left(\frac{V}{100{\,\rm km\,s^{-1}}}\right)^{-8}\nonumber\\
    &&\times\left(\frac{v_{\rm esc}}{100\,\rm km\,s^{-1}}\right)^{2}\left(\frac{a_{\rm bin}}{5R_*}\right)^{-2}\left(\frac{R_*}{500\,\Rsun}\right)^{-2}\left(\frac{t}{t_{\bullet}}\right)^{-2}\ .
    \label{eq:r_disk}
\end{eqnarray}
Hence an accretion disk can form outside the CO if $r_{\rm disk}>r_{\rm in}$, where we adopt $r_{\rm in}=6GM_{\bullet}/c^2\sim 9\times 10^6\ {\rm cm}\ (M_\bullet/10M_\odot)$. The accretion rate for $t_{\bullet}<t<a_{\rm bin}/v_{\rm min}$ is given by the Bondi-Hoyle-Lyttleton rate as \citep{Hoyle&Lyttleton1939,Bondi&Hoyle1944,Bondi1952}
\begin{eqnarray}
    \dot{M}_{\rm acc}&\approx& 4\pi r_{\rm B}^2\rho_{\rm CSM}V\\
    &\sim&\frac{10^4\xi^3\dot{M}_{\rm Edd}}{\xi^3-\chi^3}\left(\frac{M_{\bullet}}{10\ \Msun}\right)\left(\frac{M_{\rm CSM}}{0.1\,\Msun}\right)\nonumber \\
    &&\times\left(\frac{V}{100\ {\rm km\ s^{-1}}}\right)^{-3}\left(\frac{a_{\rm bin}}{5R_*}\right)^{-3}\left(\frac{R_*}{500\,\Rsun}\right)^{-3}\left(\frac{t}{t_{\bullet}}\right)^{-3}\ ,\nonumber
\end{eqnarray}
where we assumed $r_{\rm B}\ll a_{\rm bin}$. Here the Eddington accretion rate is defined by $\dot{M}_{\rm Edd}\equiv \epsilon_{\rm sd}^{-1}L_{\rm Edd}/c^2 \sim 2\times 10^{19}\ {\rm g\ s^{-1}}(\epsilon_{\rm sd}/0.1)^{-1}(M_{\bullet}/10M_\odot)(\kappa/0.3\ {\rm cm^2\ g^{-1}})^{-1}$, where $\epsilon_{\rm sd}\sim 0.1$ is the radiative efficiency of a standard disk. 

Such accretion should lead to an advection-dominated accretion flow \citep[ADAF, e.g.,][]{Ichimaru1977,Narayan94}, where we expect mass outflows of high velocities \citep[e.g.,][]{Ohsuga05, Jiang14,Sadowski14,Kitaki21}. Here we use the ADAF model by \cite{Blandford&Begelman1999} to estimate the luminosity of the disk wind\footnote{We call this energy injection as ``wind" in the rest of the paper, but we note that other forms of energy injection, such as bright X-ray emission or jets, may be realized \citep[e.g.,][]{Narayan17,Soker22}.} that will energize the CSM. In this model most of the accreted material is blown away by the wind, and only a small fraction reaches the compact object. The accretion rate is assumed to be a power-law in radius (from the CO) as $\dot{M}(r)\propto r^p$ for a range $r_{\rm in}<r<r_{\rm disk}$, with $r_{\rm disk}$ defined in equation (\ref{eq:r_disk}). We adopt $p=0.5$, which is consistent with the range inferred from simulations and observations of radiatively inefficient accretion flows ($0.3 <p < 0.8$; \citealt{Yuan&Narayan2014}). \rev{The mass gain of the CO by accretion at $r_{\rm in}$ is
\begin{eqnarray}
    \Delta M_{\rm \bullet}&\sim& \left[\dot{M}_{\rm acc}\left(\frac{r_{\rm in}}{r_{\rm disk}}\right)^p\right]_{t=t_\bullet} t_\bullet \nonumber \\ 
    &\overset{p=0.5}{\sim}&\frac{10^{-5}\xi}{\xi^3-\chi^3}{\,\rm \Msun\,}\left(\frac{M_{\bullet}}{10\ \Msun}\right)\left(\frac{M_{\rm CSM}}{0.1\,\Msun}\right)\left(\frac{V}{100\ {\rm km\ s^{-1}}}\right)\nonumber\\
    &&\times\left(\frac{v_{\rm esc}}{100{\,\rm km\,s^{-1}}}\right)^{-2}\left(\frac{a_{\rm bin}}{5R_*}\right)^{-1}\left(\frac{R_*}{500\,\Rsun}\right)^{-1}\ ,\nonumber
\end{eqnarray}
which is negligible compared to $M_{\bullet}$ and hence justifies the implicit assumption of constant $M_\bullet$.}

Assuming that the wind velocity at each radius is the local escape velocity, the mass-averaged velocity of the wind is obtained from integration over radius as \citep[e.g.,][]{Metzger12, Tsuna19, Fuller22}
\begin{eqnarray}
        \langle v_{\rm wind}\rangle &\approx& \sqrt{\frac{p}{1-p}\frac{GM_\bullet}{r_{\rm in}} \left[\left(\frac{r_{\rm in}}{r_{\rm disk}}\right)^p - \left(\frac{r_{\rm in}}{r_{\rm disk}}\right) \right]} \label{eq:vwind} \\
        &\overset{r_{\rm in}\ll r_{\rm disk}}{\sim}& \sqrt{\frac{p}{1-p}\frac{GM_\bullet}{r_{\rm in}} \left(\frac{r_{\rm in}}{r_{\rm disk}}\right)^p} \nonumber \\
        &\overset{p=0.5}{\sim}& 0.04\xi^{-0.5}\,c\,\left(\frac{M_{\bullet}}{10\,~\Msun}\right)^{-1/2}\left(\frac{V}{100{\,\rm km\,s^{-1}}}\right)^{2}\nonumber\\
        &&\times \left(\frac{v_{\rm esc}}{100{\,\rm km\,s^{-1}}}\right)^{-1/2}\left(\frac{a_{\rm bin}}{5R_*}\right)^{1/2} \nonumber \\
        &&\times \left(\frac{R_*}{500\,\Rsun}\right)^{1/2} \left(\frac{t}{t_{\bullet}}\right)^{1/2}\ . \nonumber
\end{eqnarray}
The approximation $r_{\rm in}\ll r_{\rm disk}$ is used only in this section to demonstrate the approximate scaling. The wind kinetic energy luminosity is given as 
\begin{eqnarray}
    L_{\rm wind}&\approx& \frac{1}{2}\dot{M}_{\rm acc}\langle v_{\rm wind}\rangle^2 \nonumber \\
    &\overset{
    \begin{subarray}{c}
                r_{\rm in}\ll r_{\rm disk},\\
                p=0.5
    \end{subarray}
    }{\sim}&\frac{2\times 10^{41}\xi^2}{\xi^3-\chi^3}{\,\rm erg\ s^{-1}\,}\left(\frac{M_{\bullet}}{10\ \Msun}\right)\left(\frac{M_{\rm CSM}}{0.1\,\Msun}\right)\nonumber\\
    &&\times\left(\frac{V}{100\ {\rm km\ s^{-1}}}\right)\left(\frac{v_{\rm esc}}{100{\,\rm km\,s^{-1}}}\right)^{-1}\nonumber\\
    &&\left(\frac{a_{\rm bin}}{5R_*}\right)^{-2}\left(\frac{R_*}{500\,\Rsun}\right)^{-2}\left(\frac{t}{t_{\bullet}}\right)^{-2}\ .
    \label{eq:Lwind}
\end{eqnarray}
Since the wind luminosity steeply declines with time as $Vt^{-2}\propto t^{-3}$ at early times, the total available kinetic energy of the wind to power the precursor is
\begin{eqnarray}
E_{\rm wind}&\approx& \frac{1}{2} t_\bullet L_{{\rm wind},t=t_\bullet}\\
&\overset{p=0.5}{\sim}&\frac{10^{48}\xi}{\xi^3-\chi^3}{\,\rm erg\,}\left(\frac{M_{\bullet}}{10\ \Msun}\right)\left(\frac{M_{\rm CSM}}{0.1\,\Msun}\right)\left(\frac{V}{100\ {\rm km\ s^{-1}}}\right)\nonumber\\
    &&\times\left(\frac{v_{\rm esc}}{100{\,\rm km\,s^{-1}}}\right)^{-2}\left(\frac{a_{\rm bin}}{5R_*}\right)^{-1}\left(\frac{R_*}{500\,\Rsun}\right)^{-1}\ .\nonumber
\end{eqnarray}
A luminous precursor can be powered if this energy is injected to the CSM at a timescale comparable to the diffusion timescale, analogous to the central engine models proposed for super-luminous SNe \citep[e.g.,][]{Kasen&Bildsten2010,Woosley2010}.

\subsection{Transients from Wind-CSM Interaction}
The disk wind is injected to the rest of the CSM, and we assume that the kinetic energy of the injected wind quickly thermalizes in the CSM \citep[for the validity of this assumption see][]{Kimura17}. We calculate the dynamics of the CSM and the resulting light curves using the following one-zone model. 

\subsubsection{Governing Equations}
We assume that the opacity in the CSM is uniform and constant, which is justified for the characteristic density of the CSM \citep[e.g., see Figure 3 in][]{Matsumoto22b} unless significant amounts of dust are formed within the CSM. We can then solve the evolution by the following set of equations \citep[e.g., Section 2.2 of][]{Metzger15} 
\begin{eqnarray}
    \frac{dE_{\rm int}}{dt} &=&-\frac{E_{\rm int}}{R_{\rm CSM}} v_{\rm CSM} + L_{\rm inj} - L_{\rm rad} \label{eq:dEintdt}\\
    \frac{dR_{\rm CSM}}{dt} &=& v_{\rm CSM} \label{eq:drdt} \\
    \frac{dv_{\rm CSM}}{dt} &=& \frac{5}{3} \frac{E_{\rm int}}{M_{\rm CSM} R_{\rm CSM}} \label{eq:dvdt}\\
    L_{\rm rad} &=& \frac{E_{\rm int}}{t_{\rm diff}} \label{eq:L_rad}
\end{eqnarray}
where $E_{\rm int}$ is the internal energy, $L_{\rm rad}$ is the radiative luminosity, $t_{\rm diff}\approx 3\kappa M_{\rm CSM}/(4\pi c R_{\rm CSM})$ is the diffusion timescale in the CSM at time $t$, and $L_{\rm inj}$ is the rate of energy injection due to the disk wind (with $\langle v_{\rm wind}\rangle$ in equation \ref{eq:vwind})
\begin{eqnarray}
    L_{\rm inj} &=& 
  \left\{\begin{array}{@{}l@{\quad}l}
       L_{\rm wind} &  \left(t_{\bullet} < t < a_{\rm bin}/v_{\rm min}\ {\rm and}\ r_{\rm disk}>r_{\rm in}\right) \\
      0 & ({\rm otherwise})
    \end{array}\right.
    \label{eq:Linj}
\end{eqnarray}
Equation (\ref{eq:dEintdt}) is the first law of thermodynamics with PdV work in the first term, and equation (\ref{eq:dvdt}) accounts for the increase of the kinetic energy $E_{\rm kin}=3M_{\rm CSM}v_{\rm CSM}^2/10$ due to PdV work\footnote{The prefactor $3/10$ is for a homologously expanding, uniform ejecta \citep{Arnett1982}, but we note that the corresponding equation in the original paper contains a typo (3/5 should be 5/3; see also \citealt{Wheeler15}).}. As the wind carries negligible mass compared to the rest of the CSM, we can approximate $M_{\rm CSM}$ as a constant.

\subsubsection{Accounting for recombination}

Initially the CSM is ionized and the adoption of uniform electron scattering opacity for $\kappa$ is justified. However, as the CSM expands and its photospheric temperature 
\begin{eqnarray}
    T_{\rm ph}=(L_{\rm rad}/4\pi \sigma R_{\rm CSM}^2)^{1/4}\ ,
\end{eqnarray}
where $\sigma$ is the Stefan-Boltzmann constant, becomes smaller than the ionization temperature $T_{\rm I}$, recombination of the CSM sets in. Afterwards, a sharp ionization front develops and recedes within the CSM, with its radius given as $R_{\rm I} = (L_{\rm rad}/4\pi \sigma T_{\rm I}^4)^{1/2}$.
In order to calculate the light curve in this regime, the location of the ionization front should be tracked \citep{Popov1993,Dexter13,Matsumoto+2016,Matsumoto24}.

The radius of the ionization front can be parameterized by the dimensionless variable $x_i\equiv R_{\rm I}/R_{\rm CSM}(\leq 1)$, which is 1 when the CSM is fully ionized. Since the outside of $R_{\rm I}$ is transparent, Equation (\ref{eq:dEintdt}) is then modified to consider the change of internal energy in the ionized region of volume $V_i=4\pi(x_iR_{\rm CSM})^3/3$, as
\begin{eqnarray}
 \frac{1}{\epsilon_{\rm int}}\frac{\partial \epsilon_{\rm int}}{\partial t} + \frac{4}{x_iR_{\rm CSM}}\frac{\partial (x_iR_{\rm CSM})}{\partial t} = \frac{L_{\rm inj}-L_{\rm rad}}{\epsilon_{\rm int}V_i}, \label{eq:dEintdt_recomb}
\end{eqnarray}
where $\epsilon_{\rm int}=E_{\rm int}/V_i$ is the internal energy density. Assuming that the ionized CSM is still optically thick, both the Thomson optical depth and light-crossing time in the ionization region are $\propto x_i$, so the diffusion time where photons escape is modified to $x_i^2t_{\rm diff}$. The radiative luminosity is thus modified by a factor $x_i^{-2}$ as
\begin{eqnarray}
    L_{\rm rad}=\frac{\epsilon_{\rm int}V_i}{x_i^2t_{\rm diff}} = 4\pi (x_iR_{\rm CSM})^2 \left(\frac{\epsilon_{\rm int}R_{\rm CSM}}{3x_it_{\rm diff}}\right), \label{eq:Lrad_rec}
\end{eqnarray}
which is equated with the luminosity from the ionization front $L_{\rm rad}=4\pi (x_iR_{\rm CSM})^2\sigma T_I^4$ to obtain
\begin{eqnarray}
    \epsilon_{\rm int}=\frac{3x_it_{\rm diff}}{R_{\rm CSM}}\sigma T_I^4. \label{eq:eps_int}
\end{eqnarray}
Substituting this to equation (\ref{eq:dEintdt_recomb}), we obtain a differential equation for $x_i$ as
\begin{eqnarray}
    \frac{dx_i}{dt}=-\frac{2x_i}{5}\frac{v_{\rm CSM}}{R_{\rm CSM}}-\frac{1}{5x_it_{\rm diff}}\left[1-\frac{L_{\rm inj}}{4\pi (x_iR_{\rm CSM})^2\sigma T_I^4}\right]. \label{eq:dxdt}
\end{eqnarray}

Note that if the acceleration is negligible and the CSM expands freely with the relation $R_{\rm CSM}=v_{\rm CSM}t$, we recover the evolution of $x_i$ in \cite{Dexter13} (their equation A12; note that their $t_{\rm d}$ differs from our $t_{\rm diff}$ as $t_{\rm diff}=t_{\rm d}^2/t$). However the energy injection from the accreting CO can accelerate the CSM via PdV work. The PdV work in equation (\ref{eq:dvdt}) now only comes from the ionized region, but under the one-zone approximation we assume that this pushes the {\it entire} CSM with an acceleration given as
\begin{eqnarray}
    \frac{dv_{\rm CSM}}{dt} &=& \frac{5}{3}\frac{1}{M_{\rm CSM}v_{\rm CSM}}\times \frac{\epsilon_{\rm int}}{3}\frac{dV_i}{dt} \nonumber \\
    &=&\frac{4\pi (x_i R_{\rm CSM})^2\epsilon_{\rm int}}{3M_{\rm CSM}v_{\rm CSM}}\frac{d(x_iR_{\rm CSM})}{dt} \label{eq:dvdt_rec}.
\end{eqnarray}
By solving equations (\ref{eq:drdt}) and (\ref{eq:Lrad_rec}) - (\ref{eq:dvdt_rec}), we can obtain $R_{\rm CSM}, L_{\rm rad}, \epsilon_{\rm int}$, $x_i$ and $v_{\rm CSM}$ as a function of $t$ after the recombination starts. 

\subsubsection{Methods of our light curve model with recombination}
\label{sec:methods_with_recombination}

Generally, the energy injection raises the precursor luminosity and sustains the ionization of the CSM. One important difference between our situation and existing modeling \citep{Dexter13,Matsumoto24} is that the energy injection is delayed, happening only after the CSM reaches the CO. For the CSM from compact progenitors in tight binaries, this is not a significant effect because the CSM reaches the CO early and energy injection is done well before recombination sets in. In this case the formulation is the same as the modeling of normal type II-P SNe \citep{Popov1993,Kasen&Woosley2009}.

However, for RSG progenitors the CSM generally recombines before the energy injection. The former timescale is estimated by solving equation~\eqref{eq:dxdt} for $L_{\rm inj}=0$ \citep[e.g.,][]{Matsumoto+2016}:
\begin{eqnarray}
t_{\rm rec}&\approx&100\xi^{-5/14}{\,\rm day\,}\left(\frac{M_{\rm CSM}}{0.1\,\Msun}\right)^{5/14}
\nonumber\\
&&\times\left(\frac{v_{\rm esc}}{100{\,\rm km\,s^{-1}}}\right)^{-5/14}\left(\frac{R_*}{500\,\Rsun}\right)^{1/7}\ ,
\end{eqnarray}
which could be shorter than the latter one $t_\bullet\approx200\xi^{-1}\,\rm days$ (see equation~\ref{eq:t_dot}). In this case, the recombined CSM is \textit{re-}ionized by the delayed energy injection and then recombines again at late times. In our one-zone prescription, the sudden energy injection at $t=t_\bullet$ to an almost neutral ($x_i\ll 1$) CSM results in a prompt ionization, due to the last term $\propto L_{\rm inj}/x_i^3$ in equation (\ref{eq:dxdt}). While this instantaneous ionization essentially comes from our one-zone assumption, we believe that sudden re-ionization probably does occur in nature. This is because the ionization would result from shock heating following the wind-CSM collision, and this occurs over the wind-crossing timescale of the CSM $\sim R_{\rm CSM}/\langle v_{\rm wind}\rangle$ which is much faster than the dynamical timescale $\sim R_{\rm CSM}/v_{\rm CSM}$.

For energy injection into a neutral CSM ($x_i\ll 1$), the ionization will proceed as in equation (\ref{eq:dxdt}), and saturates when either $x_i$ reaches 1 (full ionization) or $dx_i/dt=0$. In the latter case, the corresponding degree of ionization is obtained by solving for $x_i$ in equation (\ref{eq:dxdt}):
\begin{eqnarray}
    x^2_{\rm sat} &=& -\frac{R_{\rm CSM}} {4v_{\rm CSM}t_{\rm diff}}\nonumber \\
    &&+\sqrt{\left(\frac{R_{\rm CSM}}{4v_{\rm CSM}t_{\rm diff}}\right)^2 + \frac{R_{\rm CSM}}{2v_{\rm CSM}t_{\rm diff}}\frac{L_{\rm inj}}{4\pi R_{\rm CSM}^2\sigma T_I^4}}\ .
    \label{eq:x_sat}
\end{eqnarray}
Then, the location of the ionization front at saturation is estimated as $x_{i, \bullet} = {\rm min}(1, x_{\rm sat})$. For a reliable calculation of the light curve, we need to smoothly evolve the ionization from the $x_i$ just before energy injection to $x_{i, \bullet}$, and also accurately evolve the equations that control the energy budget ($E_{\rm int}, E_{\rm kin}$) of the precursor.

For every time $t$, we evolve the first set of equations (\ref{eq:dEintdt}) -- (\ref{eq:L_rad}) when the CSM is fully ionized ($x_i=1$), and when $x_i<1$ ($R_{\rm I}<R_{\rm CSM}$) we evolve the second set of equations (\ref{eq:drdt}), (\ref{eq:Lrad_rec}) -- (\ref{eq:dvdt_rec}) that also follow the evolution of $x_i$. We control the timestep of the integration by the equations shown in Appendix \ref{sec:dt_control}. If the CSM becomes fully ionized during the energy injection, we switch back to solving the equations for $x_i=1$, with $E_{\rm int}=\epsilon_{\rm int}(4\pi R_{\rm CSM}^3/3)$ at that time.

For the initial conditions at $t=0$, we set $R_{\rm CSM,0}=R_*$, and assume the kinetic and internal energy are equally shared with total energy $3M_{\rm CSM}(\xi v_{\rm esc})^2/10$, i.e. $E_{\rm kin}=E_{\rm int}=3M_{\rm CSM}(\xi v_{\rm esc})^2/20$. We set a floor value for $x_i$ of $10^{-5}$, as it increasingly becomes small as the CSM recombines. We verified that enhancing the floor value to $10^{-4}$ and $10^{-3}$ results in almost identical light curves.

There are eight parameters characterizing our binary precursor model, $M_\bullet$, $M_*$, $R_*$, $\kappa$, $T_I$, $\xi$, $M_{\rm CSM}$, and $a_{\rm bin}$. For simplicity, in this work we fix $\xi=2$ as in \cite{Kimura17}, and consider two values $M_{\bullet}=\{10, 1.4\}\ M_\odot$ representing a BH and NS companion respectively. We discuss the effect of changing $\xi$ in Appendix \ref{sec:vary_xi}, where we find it to not greatly affect the light curves. For the SN progenitor, we consider a RSG and two He star progenitors, with values ($M_*, R_*, \kappa, T_I$) in Table \ref{tab:params}. We consider a high-mass He star with a compact radius that is expected to be a typical progenitor of stripped-envelope SNe, and a low-mass He star with mass around $3M_\odot$ that expands to a radius of around $\sim \! 100R_\odot$ before core-collapse \citep[e.g.,][]{Woosley19,Ertl20,Laplace2020,Wu22b}. The values of $\kappa$ and $T_I$ for the two types of progenitors are motivated from singly-ionized hydrogen and helium respectively \citep[e.g.,][]{Kleiser14}. We consider a range of values for the remaining free parameters $M_{\rm CSM}$ and $a_{\rm bin}$, which should greatly vary for each SN progenitor. For $M_{\rm CSM}$ we adopt $0.1M_\odot\leq M_{\rm CSM}\leq 3M_\odot$ for RSGs and $0.01M_\odot\leq M_{\rm CSM}\leq 1M_\odot$ for He stars (see Section \ref{sec:single_lum}). \rev{For $a_{\rm bin}$ we consider $3R_*\leq a_{\rm bin}\leq 30R_*$ for all models, with the upper limit inspired from our finding that beyond this limit most models predict precursors dimmer than those observed (Figure \ref{fig:lc_properties}).}

\section{Results}
\label{sec:result}

\subsection{Example Light Curves}
\begin{figure*}
    \centering
    \includegraphics[width=\linewidth]{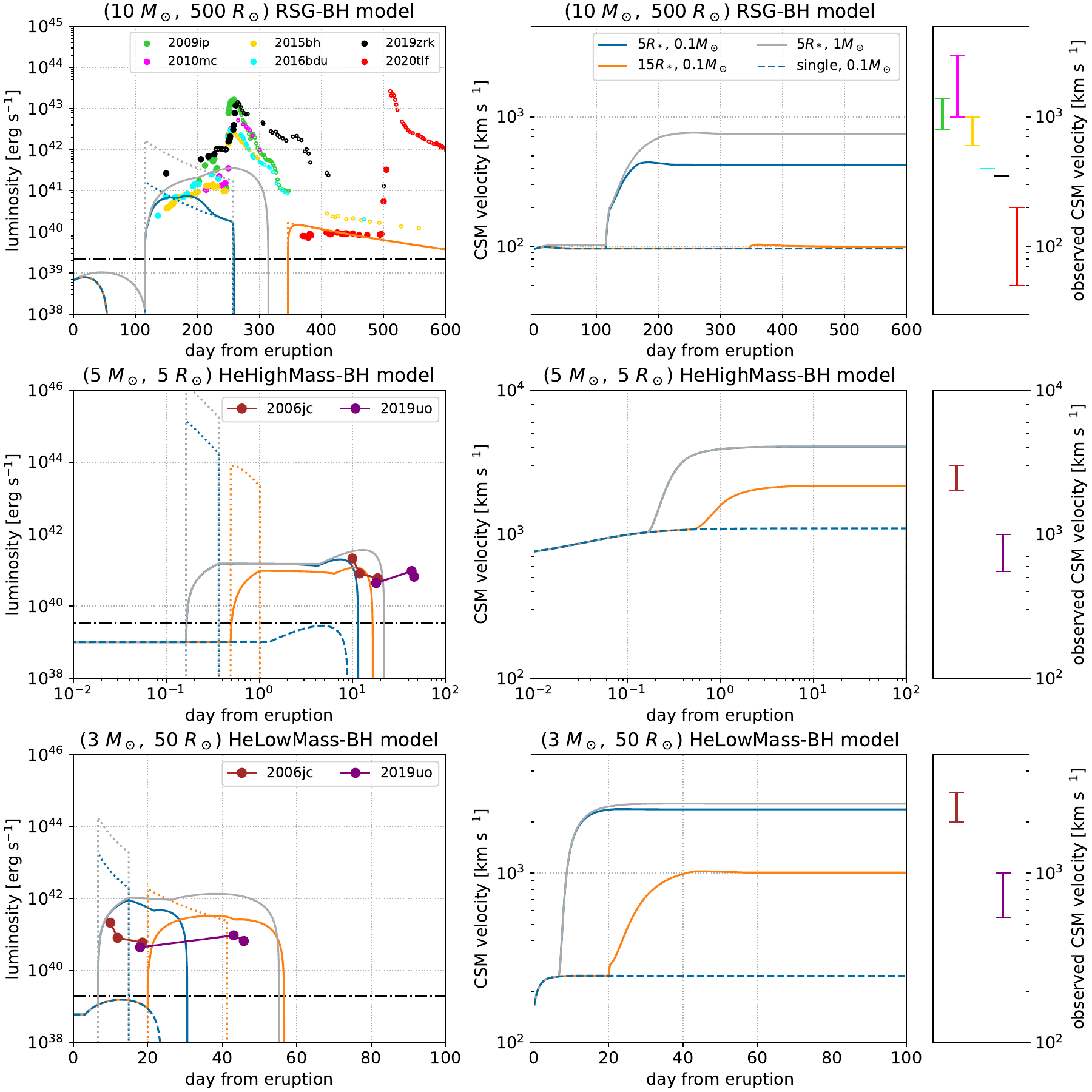}
    \caption{Light curves and CSM velocities for a few parameter sets of ($a_{\rm bin}$, $M_{\rm CSM}$), for the case of a $10M_\odot$ BH companion (other parameters are in Table \ref{tab:params}). Dotted lines show the wind (injection) luminosity $L_{\rm wind}$ for each parameter, and dashed lines show the injection-free cases, i.e., a single star progenitor.
    The points show the light curves of eight well-observed SN precursor samples: SN 2019zrk \citep{Fransson22}, 2006jc \citep{Pastorello07}, 2019uo \citep{Strotjohann21}, and others from \citealt{Matsumoto22}. For SN 2006jc and 2019uo, the precursors are observed $\approx2\,\rm yr$ and $\approx 300\,\rm day$ before the SNe, respectively.
    For the data points, the times of SN explosions are set so that the points are roughly aligned with our light curves.
    The post-peak part of the SN light curve are in small empty circles, to distinguish with the precursor emission (large filled circles) we aim to reproduce. The horizontal dash-dotted lines show our luminosity limit for the single star case (equation~\ref{eq:Lprec_single}). The right panels show the observed CSM velocities of these events by \citet{Strotjohann21} (SN 2019zrk and SN 2019uo); \cite{Foley07} (SN 2006jc); \cite{Matsumoto22} and references therein (others). Note that in the center panel the blue line is overlapped with the gray line; while larger $M_{\rm CSM}$ gives larger $L_{\rm wind}$, the increased inertia results in an acceleration independent of $M_{\rm CSM}$.
    }
    \label{fig:few_cases_noepswind_BH}
\end{figure*}
\begin{figure*}
    \centering
    \includegraphics[width=\linewidth]{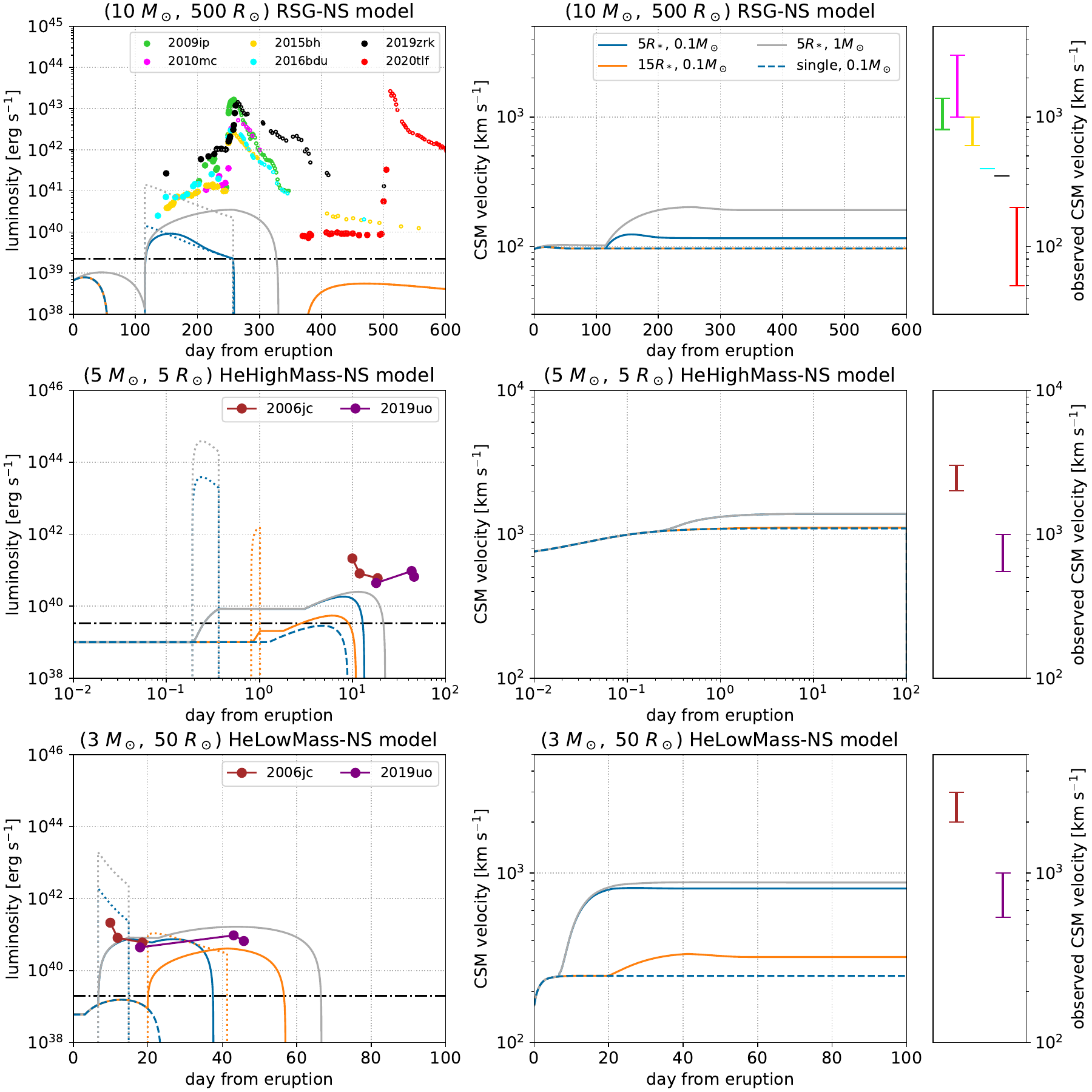}
    \caption{Same as Figure \ref{fig:few_cases_noepswind_BH}, but for the case of a $1.4M_\odot$ NS companion.}
    \label{fig:few_cases_noepswind_NS}
\end{figure*}

Figure~\ref{fig:few_cases_noepswind_BH} depicts the evolution of luminosities and the CSM velocities for different parameter sets and progenitors, in the case of a BH companion of $M_\bullet=10\,\Msun$. The left panels show the light curves (solid curves) and corresponding wind (injection) luminosity $L_{\rm wind}$ (dotted curves). For RSG-BH binaries (top panel) and our fiducial parameters of $M_{\rm CSM}=0.1\,\Msun$ and $a_{\rm bin}=5\,R_*(\approx4\times10^{14}\,\rm cm)$, the CSM completely recombines $\approx 50\,\rm days$ after the ejection. The CSM starts to accrete onto the CO at $t_\bullet\approx100\,\rm days$, which triggers a powerful disk wind and energy injection until $\approx250\,\rm days$. Since the diffusion timescale is almost comparable to the energy injection time $t_\bullet$, the resulting light curve traces the injected $L_{\rm wind}$. Therefore, the light curve shows two distinct components: an early injection-free part and a later part due to energy injection. It should be noted that the former corresponds to the precursor emission from a single star, and its luminosity $\sim \! 10^{39}\,\rm erg\,s^{-1}$ is consistent with our luminosity limit in equation~\eqref{eq:Lprec_single}. As shown in the middle panel, once the energy injection starts the CSM is accelerated until recombination sets in.

Inspecting the dependence of the light curve on the parameters $a_{\rm bin}$ and $M_{\rm CSM}$, a more massive CSM (gray curve) increases the diffusion time (equation~\ref{eq:t_prec0}) and wind luminosity (equation~\ref{eq:Lwind}). Since $t_\bullet$ is independent of the CSM mass, the longer diffusion time makes (i) the light curve broader and deviate from $L_{\rm wind}$, and (ii) delays the recombination which results in higher CSM velocity. A larger binary separation (orange curve) simply delays the energy injection, which leads to smaller $L_{\rm wind}$ and makes the light curve almost identical to the evolution of $L_{\rm wind}$. Therefore, in this regime the duration and luminosity of the precursor are given by
\begin{align}
t_{\rm prec,1}&\approx\frac{a_{\rm bin}}{v_{\rm min}} - \frac{a_{\rm bin}}{v_{\rm max}} \nonumber\\
&\sim 200{\,\rm day\,}\left(\frac{1}{\chi}-\frac{1}{\xi}\right)\left(\frac{v_{\rm esc}}{100\,\rm km\,s^{-1}}\right)^{-1}
    \nonumber\\
&\times \left(\frac{a_{\rm bin}}{5R_*}\right)\left(\frac{R_*}{500\,\Rsun}\right)\ ,
    \label{eq:t_prec1}\\
L_{\rm prec,1}&\approx L_{\rm wind}
    \nonumber\\
&\sim2\times 10^{41}{\,\rm erg\ s^{-1}\,}\left(\frac{\xi^3}{\xi^3-\chi^3}\right)\left(\frac{M_{\bullet}}{10\ \Msun}\right)
    \nonumber\\
&\times\left(\frac{M_{\rm CSM}}{0.1\,\Msun}\right)\left(\frac{a_{\rm bin}}{5R_*}\right)^{-2}\left(\frac{R_*}{500\,\Rsun}\right)^{-2}\ ,
    \label{eq:L_prec1}
\end{align}
where we used $V=\xi v_{\rm esc}$ and $t=t_\bullet$ to estimtate the luminosity.

For the compact He star models, energy injection generally happens before recombination in the CSM completes. In particular for the high-mass He star models (middle panels) the injection timescale of $t_{\bullet}\approx 0.1-1\,\rm day$ is shorter than the diffusion timescale of $t_{\rm prec,0}\approx10\,\rm day$. Therefore the evolution is practically the same as the CSM ejected with an initial radius of $a_{\rm bin}$ and energy of $E_{\rm wind}$. Before the recombination, the luminosity is given by
\begin{align}
L&\approx1\times10^{41}{\,\rm erg\,s^{-1}\,}\left(\frac{\xi^2}{\xi^3-\chi^3}\right)
    \nonumber\\
&\times\left(\frac{M_{\bullet}}{10\,\Msun}\right)\left(\frac{v_{\rm esc}}{1000\,\rm km\,s^{-1}}\right)^{-1}\ ,    
\end{align}
where we used equation~\eqref{eq:Lprec_single} but replacing $E_{\rm kin}$, $R_*$, and $v_{\rm esc}$ with $E_{\rm wind}$, $a_{\rm bin}$, and $v_{\rm acc}\equiv\sqrt{10E_{\rm wind}/3M_{\rm CSM}}$ (the CSM velocity after the acceleration), respectively. Note that the normalization of the escape velocity is different from the case of RSG progenitors.
Interestingly the luminosity is independent of the CSM mass and binary separation, as roughly seen in our results. By using this luminosity, the onset time of the recombination is estimated by
\begin{align}
t_{\rm i}&\equiv\left(\frac{L}{4\pi v_{\rm acc}^2\sigma T_{\rm I}^4}\right)^{1/2}\approx6{\,\rm day\,}\left(\frac{a_{\rm bin}}{5R_*}\right)^{1/2}\left(\frac{R_*}{5\,\Rsun}\right)^{1/2}\ .
\end{align}
During the recombination, the light curve is described by the Popov model \citep{Popov1993}, which gives the duration and luminosity \citep{Matsumoto+2016,Matsumoto24}
\begin{align}
t_{\rm prec,2}&\approx7^{5/14}t_{\rm i}^{2/7}t_{\rm prec,0}^{5/7}
    \nonumber\\
&\sim20{\,\rm day\,}\left(\frac{\xi^2}{\xi^3-\chi^3}\right)^{-\frac{5}{28}}\left(\frac{M_{\bullet}}{10\,\Msun}\right)^{-\frac{5}{28}}\left(\frac{M_{\rm CSM}}{0.1\,\Msun}\right)^{\frac{5}{14}}
    \nonumber\\
&\times\left(\frac{v_{\rm esc}}{10^3\,\rm km\,s^{-1}}\right)^{\frac{5}{28}}\left(\frac{a_{\rm bin}}{5R_*}\right)^{\frac{9}{28}}\left(\frac{R_*}{5\,\Rsun}\right)^{\frac{9}{28}}\ ,
    \label{eq:t_prec2}\\
L_{\rm prec,2}&\approx\frac{7^{3/7}56}{55}\pi \sigma T_{\rm I}^4v_{\rm acc}^2t_{\rm i}^{8/7}t_{\rm prec,0}^{6/7}
    \nonumber\\
&\sim1\times10^{41}{\,\rm erg\,s^{-1}\,}\left(\frac{\xi^2}{\xi^3-\chi^3}\right)^{\frac{11}{14}}\left(\frac{M_{\bullet}}{10\,\Msun}\right)^{\frac{11}{14}}
    \nonumber\\
&\times\left(\frac{M_{\rm CSM}}{0.1\,\Msun}\right)^{\frac{3}{7}}\left(\frac{v_{\rm esc}}{10^3\,\rm km\,s^{-1}}\right)^{-\frac{11}{14}} \nonumber \\
&\times \left(\frac{a_{\rm bin}}{5R_*}\right)^{-\frac{3}{14}}\left(\frac{R_*}{5\,\Rsun}\right)^{-\frac{3}{14}}\ .
    \label{eq:L_prec2}
\end{align}
Here, again we replaced $v_{\rm CSM}$ in $t_{\rm prec,0}$ with $v_{\rm acc}$. In particular, the duration weakly depends on the parameters.

When recombination sets in (seen as the small kinks in the light curves), the light curve temporarily rises and then steeply drops. This is due to the rapid recession of the ionization front, which quickly reduces the diffusion time and initially enhances the luminosity. We do not expect that this will always happen in reality, as the detailed evolution of the ionization front will depend on the density profile of the CSM.

A sample of six well-observed precursors of SN II \citep[taken from][]{Matsumoto22,Fransson22} and two precursors of SN Ibn \citep{Pastorello07,Strotjohann21} are plotted as colored dots. For the case of a BH companion, the range of luminosity observed in these precursors ($\sim \! \! 10^{40}$--$10^{42}$ erg s$^{-1}$) is well reproduced by the range in binary separation, given a CSM mass typical for these SNe. In particular, we find that qualitative properties of these precursor light curves could be dictated by two timescales, the beginning of energy injection $t_\bullet$ and diffusion timescale $t_{\rm prec,0}$ (see below for a more quantitative discussion). While the smooth light curve of SN 2020tlf may indicate $t_{\rm prec,0}\lesssim t_\bullet$, the other precursors have longer diffusion timescales. However, we note that the detailed morphology of the light curve can depend on the CSM profile, especially when the diffusion timescale in the CSM is shorter than the wind duration. Thus we do not attempt to make a detailed fit of the light curves here, and leave this to a future numerical study that considers a realistic density/velocity profile of the ejected CSM.

Figure~\ref{fig:few_cases_noepswind_NS} depicts the results for the NS companion case, where the same calculations are done but for a reduced $M_\bullet=1.4~M_\odot$. While the qualitative properties are the same as the BH case, the precursor is dimmer and CSM acceleration is weaker due to the smaller Bondi accretion rate. In this case, the low-mass He star model can reproduce the observed luminosity of SN Ibn precursors.

\rev{Comparison of our model to the observed precursors for both SN IIn and Ibn shows that the eruption of the envelope should have occurred around a year to a few years prior to core-collapse. For typical massive stars of $10\Msun\lesssim M_*\lesssim 30\Msun$ this corresponds to the oxygen/neon burning stages \citep{Woosley02}, which may be more promising than the other burning stages for some scenarios of envelope ejection \citep[e.g.,][]{Wu&Fuller2021}.}

\subsection{Parameter Exploration of Light Curves}
\label{sec:param_exploration}

We next conduct a parameter study to understand the expected range of light curve properties by the model parameters ($M_{\rm CSM}$, $a_{\rm bin}$). We logarithmically sample the range of $M_{\rm CSM}$ and $a_{\rm bin}$ in Table \ref{tab:params} by ten points, resulting in a total of 100 parameter sets for each progenitor model. We focus on the light curve from $t=t_\bullet$, and define the precursor duration as the timescale over which $10$\% to $90$\% of the total radiated energy is radiated away. In other words, we solve for $t_{\rm 10}, t_{\rm 90}$ from
\begin{equation}
\int_{t_{\rm 10}}^{\infty}L_{\rm rad}dt = \int_{t_\bullet}^{t_{\rm 90}}L_{\rm rad}dt = 0.9\int_{t_\bullet}^{\infty}L_{\rm rad}dt,
\end{equation}
and then define the duration and luminosity of the precursor respectively as 
\begin{eqnarray}
    t_{\rm prec}=t_{\rm 90}-t_{\rm 10}, \ L_{\rm prec}=\frac{1}{t_{\rm 90}-t_{\rm 10}}\int_{t_{\rm 10}}^{t_{\rm 90}}L_{\rm rad}dt.
\end{eqnarray}

\begin{figure*}
    \centering
    \includegraphics[width=\linewidth]{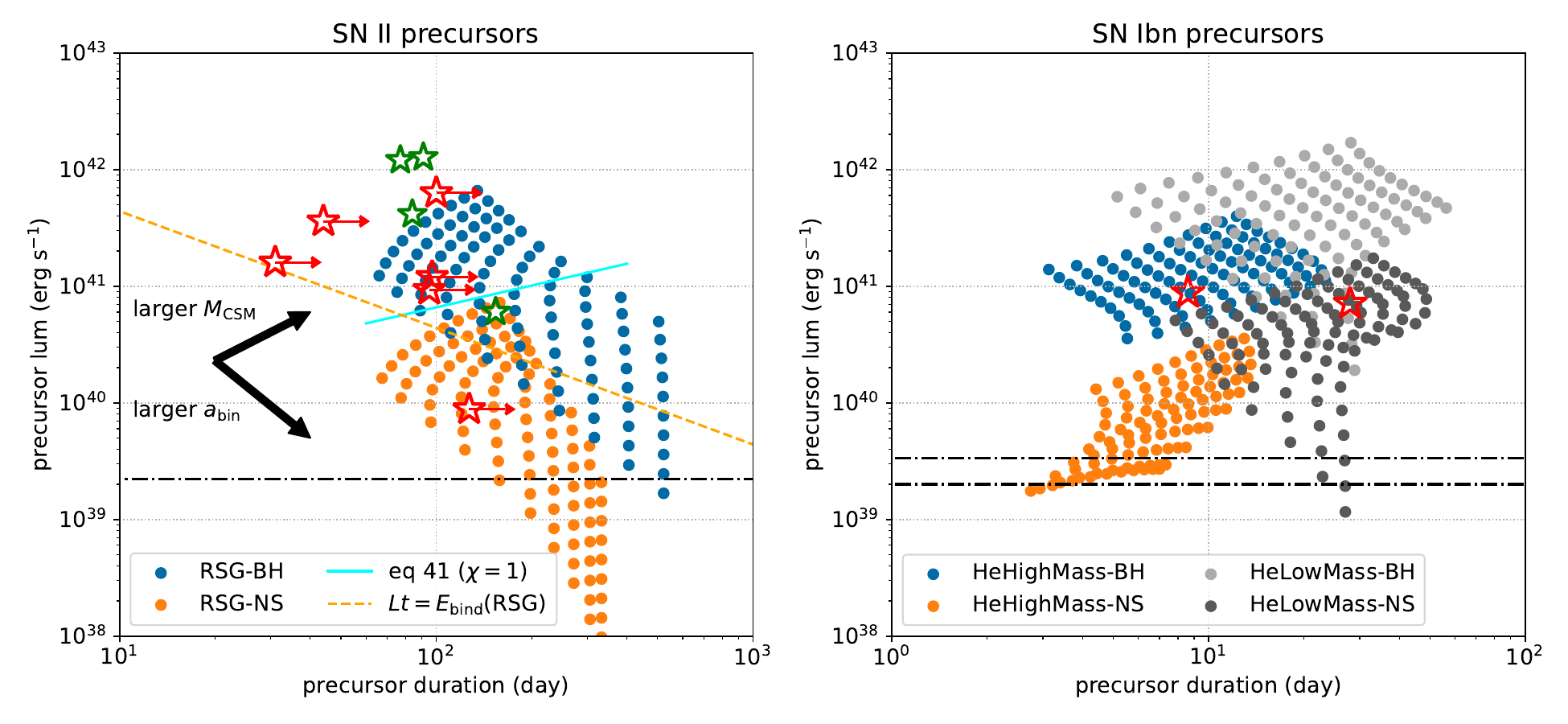}
    \caption{Light curve properties of our precursor models (dots), compared with the observed well-sampled precursors in Figure \ref{fig:few_cases_noepswind_BH} (red stars) and selected well-sampled events of \cite{Strotjohann21} with 5 or more unbinned detections in multiple bands (green stars). The SN IIn precursors have been shining until core collapse, so their durations are lower limits. The dashed orange line is the binding energy (equation \ref{eq:Ebind}) for the RSG model, and the horizontal dash-dotted lines are the luminosity limit for single stars (equation \ref{eq:Lprec_single}). The solid blue line in the left panel represents the boundary on which $t_{\rm prec,0}=t_{\bullet}$ holds (equation~\ref{eq:t_prec_eq_t_bullet}). Below (above) the line, the light curve is described by the wind luminosity (Popov model; \citealt{Popov1993}).} 
    \label{fig:lc_properties}
\end{figure*}

We show the dependence of these light curve properties on ($M_{\rm CSM}$, $a_{\rm bin}$) in Figure \ref{fig:lc_properties}. For the RSG-BH models, a wide range of luminosity ($10^{39}$--$10^{42}$ erg s$^{-1}$) and timescale (month to years) are found with varying $M_{\rm CSM}$ and $a_{\rm bin}$. 
For a fixed CSM mass, increasing $a_{\rm bin}$ always makes the precursors dimmer and longer. This is because the energy injection is delayed, and the more tenuous CSM gives a lower Bondi accretion rate. 

On the other hand, increasing $M_{\rm CSM}$ basically results in brighter and longer precursors, except for large $a_{\rm bin}$ where the duration becomes nearly independent of $M_{\rm CSM}$. This behavior is understood by comparing the diffusion and energy injection timescales. As we discussed in the previous section, for smaller $a_{\rm bin}$, the energy injection happens before the radiative diffusion plays a role, and the light curve is practically the same as that described by \cite{Popov1993}: $t_{\rm prec}\propto M_{\rm CSM}^{5/14}a_{\rm bin}^{9/28}$ (equation~\ref{eq:t_prec2}) and $L_{\rm prec}\propto M_{\rm CSM}^{3/7}a_{\rm bin}^{-3/14}$ (equation~\ref{eq:L_prec2}). 
In the opposite limit, the precursor basically traces the wind luminosity and hence we have $t_{\rm prec}\propto a_{\rm bin}$ (equation~\ref{eq:t_prec1}) independent of $M_{\rm CSM}$, and $L_{\rm prec}\propto M_{\rm CSM}a_{\rm bin}^{-2}$ (equation~\ref{eq:L_prec1}). The boundary between these two limits can be obtained by comparing $t_{\rm prec,0}$ and $a_{\rm bin}/\xi v_{\rm esc}$:
\begin{align}
\frac{a_{\rm bin}}{R_*} \left(\frac{M_{\rm CSM}}{0.1\,\Msun}\right)^{-1/2}\sim2\xi^{1/2}\left(\frac{R_*}{500\,\Rsun}\right)^{-1}\left(\frac{v_{\rm esc}}{100{\,\rm km\,s^{-1}}}\right)^{1/2}\ .
\end{align} 
Combining this relation with equations~\eqref{eq:t_prec2} and \eqref{eq:L_prec2}, the boundary on which $t_{\rm prec,0}=t_\bullet$ holds, is given by
\begin{align}
L_{\rm prec}&\sim2\times10^{41}{\,\rm erg\,s^{-1}\,}\left(\frac{t_{\rm prec}}{100\,\rm day}\right)^{18/29}\left[\frac{\xi^{81/116}}{(\xi^3-\chi^3)^{26/29}}\right]
	\nonumber\\
&\times\left(\frac{M_{\bullet}}{10\,\Msun}\right)^{{26}/{29}}\left(\frac{v_{\rm esc}}{100\,\rm km\,s^{-1}}\right)^{-{32}/{29}}\left(\frac{T_{\rm I}}{6000\,\rm K}\right)^{60/29}\ ,
\label{eq:t_prec_eq_t_bullet}
\end{align}
where we show the dependence on the recombination temperature explicitly. We plot this boundary in Figure~\ref{fig:lc_properties}, which roughly captures the transition of two cases.

For the case of a NS companion, the resulting duration and luminosity are basically the same as the BH case, but the luminosity is reduced by an order of magnitude (note $L_{\rm prec,1}\propto M_\bullet$, and $L_{\rm prec,2}\propto M_\bullet^{11/14}$). At large $a_{\rm bin}$ where the radiated luminosity traces the wind luminosity, the duration only weakly increases as $a_{\rm bin}$ increases. This is because the accretion disk formation is delayed for a NS companion until the CSM velocity at the Bondi radius, $V\approx a_{\rm bin}/t$, becomes small enough so that $r_{\rm disk}>r_{\rm in}$ is satisfied.

The He star models generally predict shorter durations of weeks to months, with the high mass models generally resulting in dimmer and faster light curves when the parameters are fixed. This is because for a fixed $a_{\rm bin}/R_*$, the high-mass models with smallest $R_*$ have the shortest accretion time scales, and lose most of the injected internal energy into adiabatic expansion before it can be radiated.

In Figure \ref{fig:lc_properties} we also show a more detailed comparison with the observed Type IIn and Ibn precursors in the luminosity-duration phase space. We note that the observed duration gives only a lower limit on the possible intrinsic duration, because when the SN explosion occurs, precursors are outshined by the SN radiation. While the RSG-BH models reproduce the range of precursor luminosities, the RSG-NS models can reproduce only the dimmer precursors of $L_{\rm prec}\lesssim 10^{41}$ erg s$^{-1}$. The parameters of the precursors for the He star models are found to cluster in a narrower range of $10^{40}$--$10^{42}$ erg s$^{-1}$, with timescales from weeks to months. These roughly overlap with the two precursors observed for type Ibn SNe, with SN 2006jc (SN 2019uo) better reproduced by the HeHighMass-BH (HeLowMass-NS) model. 

\subsection{Velocity of Accelerated CSM}
\begin{figure*}
   \centering
    \begin{tabular}{c}
     \begin{minipage}[t]{0.5\hsize}
    \centering
 \includegraphics[width=\linewidth]{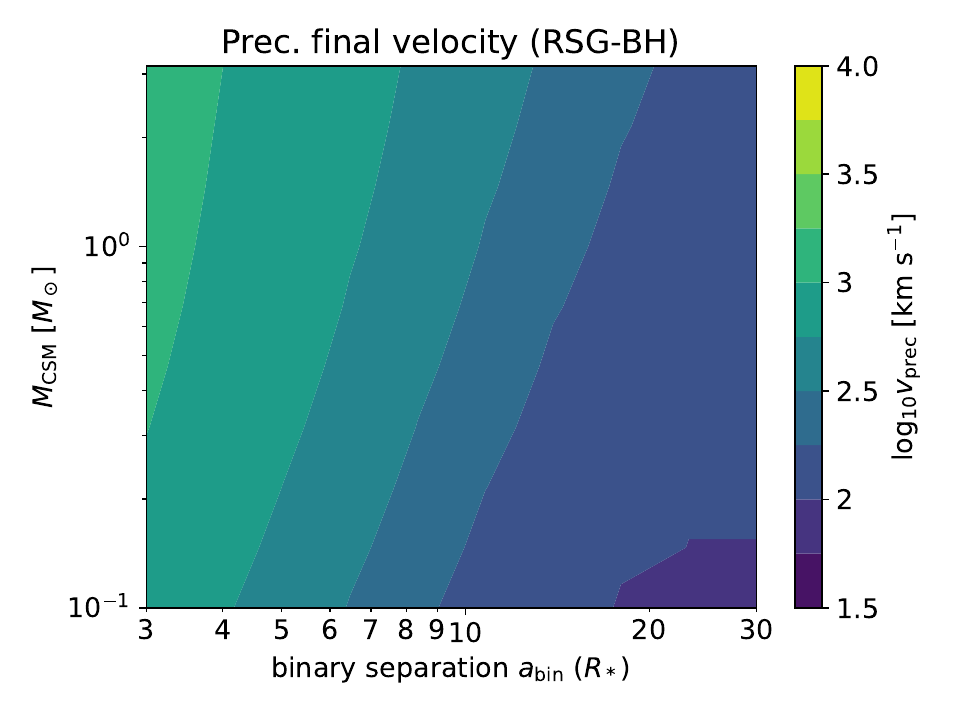}
     \end{minipage}
     \begin{minipage}[t]{0.5\hsize}
    \centering
 \includegraphics[width=\linewidth]{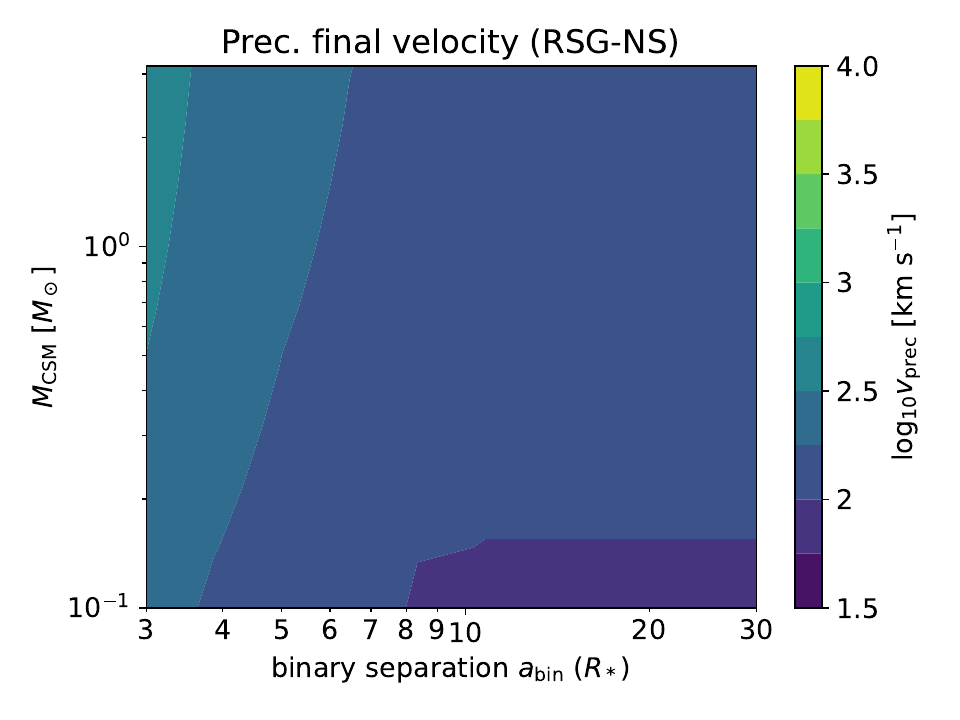}
     \end{minipage}\\
     
     \begin{minipage}[t]{0.5\hsize}
    \centering
\includegraphics[width=\linewidth]{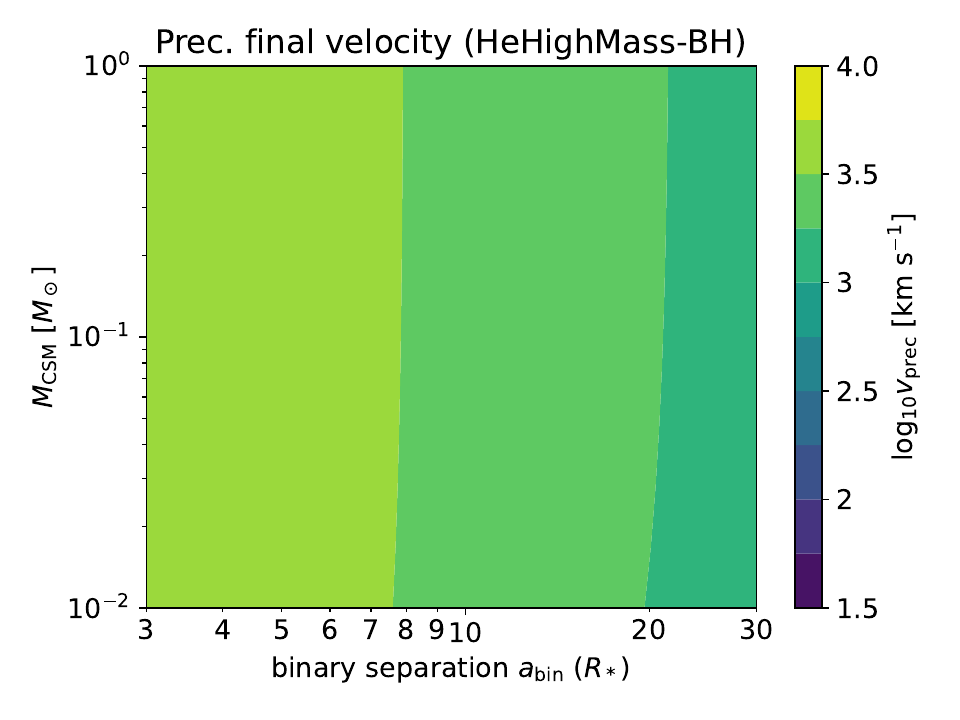}
     \end{minipage} 
     \begin{minipage}[t]{0.5\hsize}
    \centering
\includegraphics[width=\linewidth]{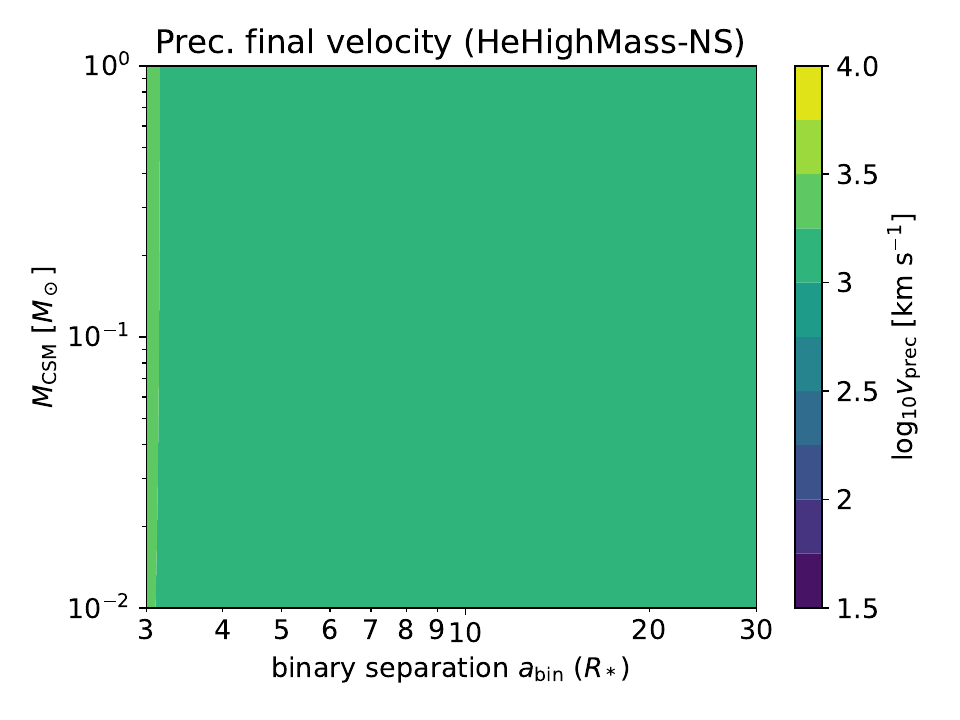}
     \end{minipage} \\
     
      \begin{minipage}[t]{0.5\hsize}
    \centering
\includegraphics[width=\linewidth]{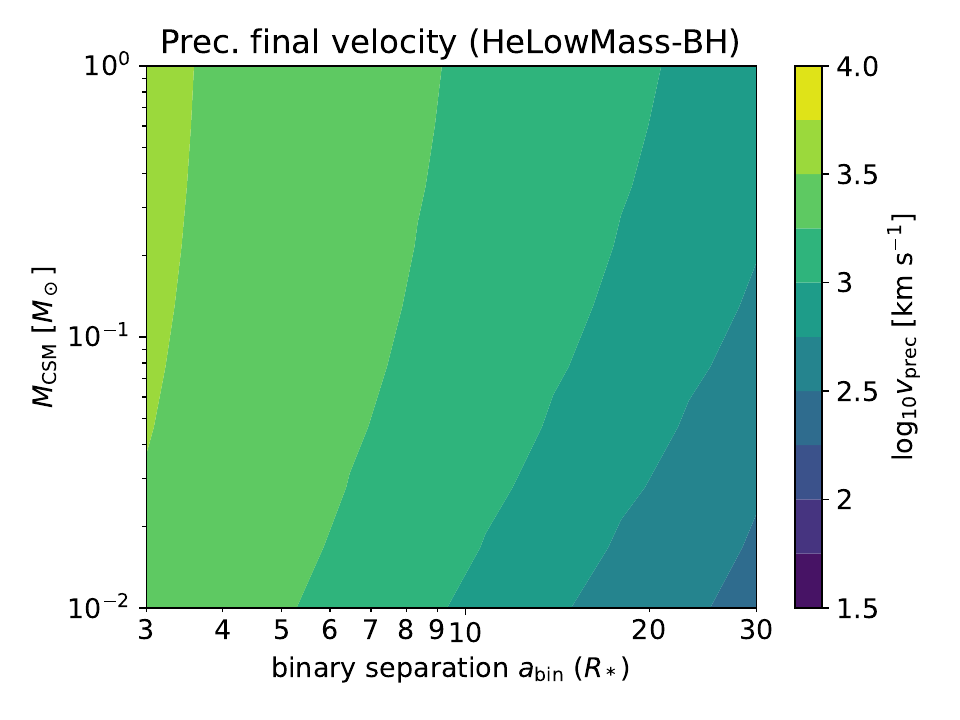}
     \end{minipage} 
     \begin{minipage}[t]{0.5\hsize}
    \centering
\includegraphics[width=\linewidth]{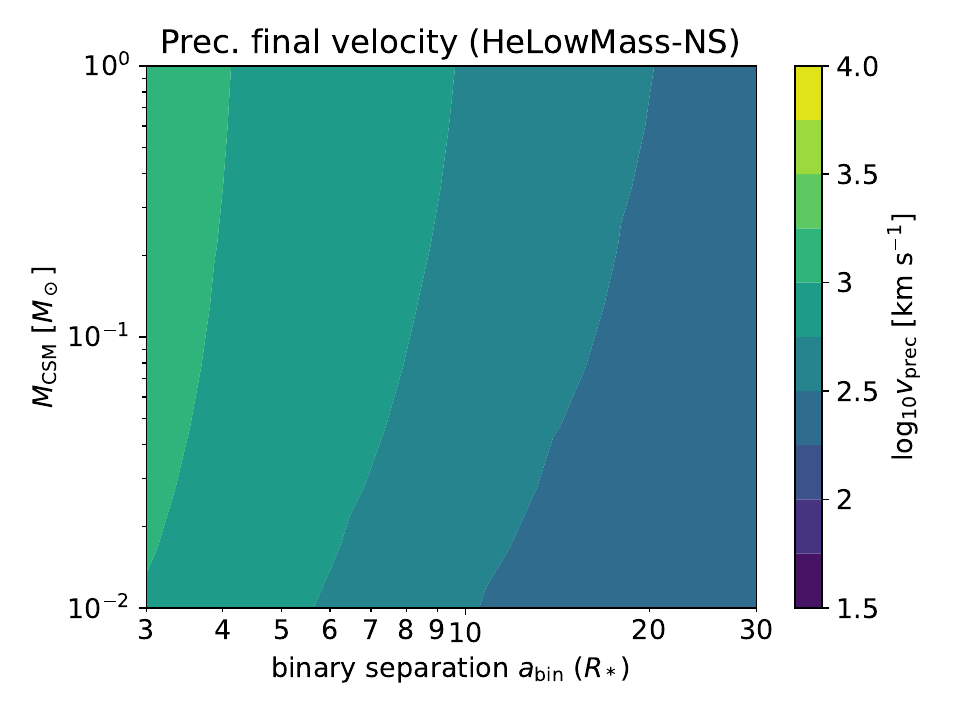}
     \end{minipage}
    \end{tabular}
\caption{Final velocity of the precursor CSM as a function of $a_{\rm bin}$ and $M_{\rm CSM}$. The HeHighMass-NS model has a negligibly small acceleration for the entire parameter space (see also Figure \ref{fig:few_cases_noepswind_NS}), while other models predict larger acceleration for smaller $a_{\rm bin}$ and larger $M_{\rm CSM}$.}
\label{fig:2Dplot_v}
\end{figure*}

Figure \ref{fig:2Dplot_v} shows the final velocity of the precursor ejecta as a function of $(M_{\rm CSM}, a_{\rm bin})$. For the RSG-BH models, the final velocities generally agree with CSM velocities observed in SN IIn, from 100 km s$^{-1}$ to $\sim \! 10^3$ km s$^{-1}$ \citep{Kiewe+2012,Taddia+2013}. It is noteworthy that this is reproduced even for RSG progenitors which are generally expected to have slow CSM velocities. 

However, a direct comparison to the observed CSM velocities on an event-by-event basis is not straightforward. For instance, broad emission lines of $1000-3000$ km s$^{-1}$ are observed in SN 2010mc, but broad wings are not solely due to fast CSM, and can also be created by Thomson scattering in the CSM if the CSM is optically thick \citep[e.g.,][]{Chugai01,Huang2018,Ishii24}. Furthermore, for the CSM velocities measured after explosion (2010mc, 2016bdu, 2019zrk, 2020tlf), they have additional contributions from radiative acceleration by the bright SN emission \citep{Chevalier&Irwin2011,Kochanek2019,Tsuna23_CSMacc}.

For SN 2009ip, multiple velocity features are observed {\it during the outburst}, from $800$--$1400$ km s$^{-1}$ typical for CSM in SN IIn, but also with a broad $\sim \! 10^4$ km s$^{-1}$ absorption feature \citep{Pastorello+2013,Margutti+2014}. Similar coexistence of slow components ($600$--$1000$ km s$^{-1}$) and fast emission line wings ($2600$--$6000$ km s$^{-1}$) are observed in SN 2015bh during the outburst phase \citep{EliasRosa+2016}. While the slow component can be explained by our model, the fast part cannot. It could be explained if we are directly observing the accretion disk wind as well as the CSM \citep[equation \ref{eq:vwind}; see also][]{Tsebrenko13}. Multi-dimensional simulations would help understand the observability of such fast outflows.

For SN Ibn, the observed CSM velocities are typically of the order of $\sim \! 10^3$ km s$^{-1}$. For compact He stars this can be easily achieved, as the CSM can be close to this velocity even without energy injection. On the other hand, for the low-mass extended He star models, significant energy injection is needed to accelerate the CSM to the observed values. This would require either a small orbital separation or a BH companion, but the latter may be in tension with observations for some SN Ibn, as it predicts bright precursors (Figure \ref{fig:lc_properties}). Radiative acceleration may also bring the CSM velocity of the low-mass He star models closer to the observed values. Due to its slower velocity, the precursor CSM for these models is likely to be more confined at the time of SN, and will receive significant acceleration by the subsequent SN radiation (e.g., Figure 6 of \citealt{Tsuna23_CSMacc}).

While the HeHighMass-BH and HeLowMass-NS models predict similar precursor luminosities that are consistent with the observed ones, the final velocities of the CSM are quite different. As shown in Figure \ref{fig:2Dplot_v}, the former would lead to CSM velocities of a few $1000$ km s$^{-1}$, while the latter would have velocities less than 1000 km s$^{-1}$ for the most of the parameter space. These are consistent with the measured velocities of SN 2006jc (2000--3000 km s$^{-1}$; \citealt{Foley07}) and SN 2019uo (550--1000 km s$^{-1}$; \citealt{Strotjohann21}) respectively. This implies that the precursors of SN Ibn may be a mixture of two binary populations, low-mass (a few $M_\odot$) He stars with NS companions and more massive He stars with BH companions.

\section{Discussion}
\label{sec:discussion}
\subsection{Possible Formation Channels and Rates}
So far we have discussed a phenomenological model of a mass ejection in close binaries, and demonstrated that they can reproduce the energetics of observed precursors. Here we outline the binary channels that produce these systems, and roughly estimate their rates. 

Our channel requires a massive star in a binary with a compact object, at a separation of $\sim \!$ 3--30$~R_*$ at the time of core-collapse. Population synthesis predicts that the outcome of massive binary systems are nearly equally split among stellar merger, envelope stripping and non-interacting binaries \citep{Sana12}. For the latter two channels, the binary can become unbound following the core-collapse of the first object, due to the natal kick of the CO. While the fraction is very sensitive to the magnitude of natal kicks, population synthesis calculations find a binary survival fraction of 14$^{+20}_{-10}$\%, with comparable amounts of NS and BH remnants \citep[][Figure 4]{Renzo19}. The surviving binaries with NS companions are weighted towards close separations and massive secondaries, due to NSs typically having larger kicks and (SN) ejecta masses than BHs. \rev{From Figure D1 in \cite{Renzo19}, the separation just before the first core-collapse has a bimodal distribution, with the first peak at a few $100~\Rsun$ (a few 10\% of the whole) and a broader second peak at $10^3$--$10^4~\Rsun$ ($\sim$ 50\%). From the orbital velocity in equation (\ref{eq:v_orbital}), we expect that surviving binaries with NSs are largely biased towards separations in the first peak, while those with BHs populate both the first and second peaks with comparable numbers.}

We therefore expect that roughly 1 -- 10\% of core-collapse SNe can arise in the presence of a CO companion. This fraction is divided into hydrogen-rich progenitors (our RSG-CO models in most cases) and stripped-envelope progenitors (our He star-CO models), which is further complicated by various binary interaction processes (and their uncertainties) preceding the second core-collapse. 

While we cannot predict the relative fractions and their expected binary separations without in-depth population synthesis calculations, we expect that for most surviving binaries the pre-SN secondary will reside close to the CO, since it is likely to expand to a similar size to that of the primary during its post-main sequence evolution. \rev{The binaries populating the first peak in separation would interact due to the expansion and (if they avoid merger) likely become He star-NS/BH binaries with separations of a few $100~\Rsun$ or less, while those in the second peak would typically become detached RSG-BH binaries with separations $\lesssim 10^4~\Rsun$, i.e. less than a few 10 $R_*$.} The separation is further favored to be close for stripped progenitors, which are usually stripped by the CO companion. Another population synthesis estimate finds that the number of stripped-envelope SNe with CO companions is $\sim \! 1$\% \citep{Zapartas17}, which is roughly consistent with this assumption.

As discussed below, the precursor rate is expected to be $\sim \! 1\%$ of core-collapse SNe for bright IIn precursors, and $\sim \! 0.1\%$ for Ibn precursors. Hence, our scenario can plausibly reproduce the observed precursors if pre-SN outbursts or enhanced pre-SN mass loss is relatively common, with a fraction of at least $\gtrsim 10$\%. For RSG progenitors the common presence of confined dense CSM in Type II SNe may support this \citep{Morozova18,Forster18,Bruch21,Bruch23}, although there are alternative possibilities to explain the dense CSM (\citealt{Dessart17,Kochanek2019,Soker2021}; see also \citealt{Davies22}).

We note that for SN 2006jc, a point source was detected several years after the SN, whose spectral energy distribution is most consistent with a companion star of the SN progenitor \citep{Maund16,Sun20}. As the inferred stellar mass for the surviving companion ($\lesssim 12\ M_\odot$) rules out a luminous blue variable, these works conclude that it is indeed the progenitor that made the precursor outburst, and not the companion. If our scenario for the precursor is correct, this may imply that the progenitor of SN 2006jc was in a hierarchical triple system, with the surviving companion star as the outer tertiary.

\subsection{Implications for Other Observations}
For the RSG-BH progenitors, we may find the progenitors of these systems in \textit{Gaia} when they are in their main-sequence (MS) phase. Assuming a circular orbit, the orbital period is
\begin{eqnarray}
    P_{\rm orb} &\approx& 2\pi\sqrt\frac{a_{\rm bin}^3}{G(M_{\rm BH}+M_*)} \nonumber \\
                &\approx& 3.9\ {\rm yr}\left(\frac{a_{\rm bin}}{10^{14}\ {\rm cm}}\right)^{3/2} \left(\frac{M_{\rm BH}+M_*}{20M_\odot}\right)^{-1/2}.
\end{eqnarray}
Since these progenitors will be non-interacting before the precursor, we may expect that the orbital period has not greatly changed from the first BH formation. The required orbital separations of $a\lesssim$ a few $\times 10^{14}$ cm for powering bright precursors of $L_{\rm prec}\gtrsim 10^{41}$ erg s$^{-1}$ overlap with the binary periods up to which \textit{Gaia} can detect their astrometric motions, of $\lesssim \! 10$ years \citep[e.g.,][]{Shikauchi22,ElBadry23b}. Theoretical calculations also predict that such detached OB-BH binaries with long periods are more common than interacting ones seen as X-ray binaries \citep[e.g.,][]{Langer20,Janssens22}.

We estimate the number of such systems in the Milky Way, assuming that they explain the observed precursors of Type IIn SNe. Adopting a bright precursor fraction in SN IIn of $f_{\rm prec}=5$--$69$\% \citep{Strotjohann21}\footnote{We note that this fraction is for precursors brighter than -13 mag. If we assume RSG-NS systems are responsible for precursors dimmer than -14 mag ($\approx 10^{41}$ erg; Section \ref{sec:param_exploration}) and instead put the threshold at -14 mag, this fraction decreases by a factor of a few (Figure 9 of \citealt{Strotjohann21}).}, SN IIn fraction among core-collapse SNe of $f_{\rm IIn}=5\%$ \citep{Cold23}, MW core-collapse SN rate of $R_{\rm SN, MW}\approx 1.6$ per century \citep{Rozwadowska21}, and assumed lifetime of the MS star $t_{\rm MS}\sim 5\ {\rm Myr}$ after the companion becomes a BH, we expect 
\begin{equation}
f_{\rm prec}f_{\rm IIn} R_{\rm SN, MW} t_{\rm MS}\sim 200-2800    
\end{equation} 
such MS-BH binaries in the Milky Way. The value of $t_{\rm MS}$ is assumed to be roughly half of the typical lifetime of a BH progenitor $\sim 10$ Myr \citep[e.g.,][]{Woosley02}, but should greatly vary for each binary system.

Population synthesis models predict that there are $\sim 10^4$--$10^5$ detached BH binaries in the Milky Way that have period within $10$ years accessible by \textit{Gaia}, with a good fraction of those with longer periods of years (\citealt{Shikauchi23}; see also \citealt{Chawla22,Rodriguez23}). The mass of the luminous star is typically $\sim \! 1\ M_\odot$, with massive ones of $\gtrsim 8\ M_\odot$ comprising typically only $\sim (1$--few)\% depending on binary population synthesis models. Thus the number of RSG-BH binary systems are estimated to be of the order of 100 -- 1000, which roughly agrees with that estimated from the SN IIn precursor rate although both still contain large uncertainties. Ongoing observations by \textit{Gaia} can lead to a better constraint on our scenario, as long-period binary systems with BHs are starting to be discovered \citep{ElBadry23b,ElBadry23a,Tanikawa23,Chakrabarti23}.

For precursors from He stars that are progenitors of SNe Ibn, the event rates are less certain as we have only a few detections. For the case of HeHighMass-BH models that may explain the SN Ibn precursors, we may compare the precursor event rates to the NS-BH merger rate. For these binaries with separations of our interest, these two can be linked because the merger timescale by gravitational wave (GW) emission \citep{Peters64}
\begin{eqnarray}
    t_{\rm GW} &=& \frac{5a_{\rm bin}^4c^5}{256 G^3M_{\rm NS}M_{\rm BH}(M_{\rm NS}+M_{\rm BH})} \nonumber \\
            &\approx& 9.4\ {\rm Gyr} \left(\frac{a_{\rm bin}}{10R_\odot}\right)^4 \nonumber \\
            &&\times\left(\frac{M_{\rm NS}}{1.4\ M_\odot}\right)^{-1} \left(\frac{M_{\rm BH}}{10\ M_\odot}\right)^{-1} \left(\frac{M_{\rm NS}+M_{\rm BH}}{11.4\ M_\odot}\right)^{-1}
\end{eqnarray}
is less than a Hubble time for separations of $a_{\rm bin}\lesssim 10~R_\odot(M_{\rm BH}/10M_\odot)^{1/2}$, and for compact stellar radii $R_*\lesssim$ (a few) $R_\odot$ expected in massive He stars \citep[e.g.,][]{Yoon_Woosley_Langer10,Woosley19,Laplace2020}\footnote{\rev{For HeLowMass-NS systems that can also explain the SN Ibn precursors, the binaries typically have separations of $\gtrsim 100~R_\odot$. Thus the final NS-NS binaries are very unlikely to merge within a Hubble time unless the second-born NS receives a natal kick in a fine-tuned magnitude and direction.}}. We infer the bright precursor fraction of Type Ibn SNe to be $\sim \! 10\%$ since one precursor was detected out of 11 targets in \cite{Strotjohann21}, while they obtain an upper limit of $<31\%$ for precursors brighter than -13 mag. This translates to an event rate of $\sim \! 100\ {\rm Gpc^{-3}}\ {\rm yr^{-1}}$ or an upper limit of $\lesssim 300\ {\rm Gpc^{-3}}\ {\rm yr^{-1}}$, assuming a SN Ibn fraction of $1\%$ among all core-collapse SNe \citep{Pastorello08,Maeda22}. This event rate is similar to the NS-BH merger rate of $ 130^{+112}_{-69}\ {\rm Gpc^{-3}}\ {\rm yr^{-1}}$, inferred from the GW detections of two NS-BH mergers by the LIGO-Virgo-KAGRA collaboration \citep{Abbott21}. 

The Ibn precursor rate is uncertain, as it could be increased relative to the CO merger rate because it can operate in binaries with periods too long to merge via GWs. However, it could be suppressed relative to the CO merger rate if pre-SN outbursts are uncommon. Although we do not claim that these kinds of binaries are the sole channel for NS-BH mergers, this approximate agreement may motivate further investigations of the connection between these peculiar optical transients and formation of binary compact objects.

Finally, there is a handful of known high-mass X-ray binaries hosting BHs \citep{Remillard06}, and recently a few detached binaries with a massive O-type star and a putative BH companion have been identified (HD130298, \citealt{Mahy22}; VFTS 243, \citealt{Shenar22}). These binaries have rather tight orbits with periods of $\lesssim 10$ days, and they may lead to He star-BH binaries if the O-stars lose their hydrogen-rich envelopes through future binary interactions. 

\subsection{Possible Caveats and Future Avenues of our Model}
\label{sec:future_work}

We have restricted our model to the case of CSM created from outbursts, which has the advantage of both the CSM mass and velocity being well-defined. However, there are other mechanisms to lift material to larger radii, such as a steady-state wind \citep[e.g.,][]{Yoon10, Moriya14,Quataert+2016}, or envelope inflation through energy injection and/or intense shell burning \citep[e.g.,][]{Soker13,SmithArnett14,Fuller&Ro2018,Ouchi&Maeda2019,Wu22a,Wu22b} resulting in rapid mass transfer. This more gradual mass expulsion from the star can result in a steady-state energy injection from the CO, and lead to precursors more long-lasting than what we have simulated here. It would be useful future work to extend our precursor model to such cases.

\rev{In this work we have not considered interacting binaries that are undergoing Roche-Lobe overflow. If the mass transfer rate is sufficiently high, both the matter undergoing mass transfer and the circum-binary matter can complicate the accretion onto the CO and the dynamics of the precursor-generated CSM. The precursor CSM will collide with this pre-existing matter, which may make the precursor even brighter. Even in cases where the pre-existing matter has much less mass than the precursor CSM and are dynamically unimportant, the precursor is still brighter due to the tighter separation. For example, the HeLowMass-NS model with $M_{\rm CSM}=0.1~\Msun$ and $a_{\rm bin}=2.2R_*$ (approximately the Roche limit) predicts a precursor $\sim 40\%$ brighter than that with the same $M_{\rm CSM}$ and $a_{\rm bin}=3R_*$.}

Two major multi-dimensional effects are not captured in our one-zone light curve modeling. First, as energy injection starts soon after the fastest end of the CSM reaches the CO, the CSM is initially heated not from the center but from the outside. This may affect the initial evolution of the CSM, while the effect may be smaller at late times. Second, we do not consider the detailed angular dependence of the energy injection due to disk wind. In our work, the angle subtended by the wind is somewhat idealized like \cite{Kimura17}, i.e. wide enough so that it can heat the entire CSM, but not too wide such that feedback would prevent the CSM from accreting onto the CO. These effects are important caveats that should be investigated in multi-dimensional hydrodynamical simulations.

Related to these multi-dimensional effects, if a powerful disk wind or a relativistic jet hypothetically launched from the CO emerges from the CSM, it may produce radio/X-ray precursors of SN explosions. This is analogous to a model where accretion of a SN ejecta by a very close CO companion can potentially power a gamma-ray burst jet and a high-energy transient \citep[e.g.,][]{Rueda12, Fryer14}. Detecting such signals would be a smoking gun signature of our model. We plan to study their detectability in future work.

\rev{We have focused on the putative progenitors of Type IIn/II-P and Ibn SNe where precursors were observed, but there is a diversity in SNe and the structures of their progenitors. An example we have not considered is Type IIb SNe, whose progenitors are believed to have a low-mass hydrogen layer of $\sim 0.1~M_\odot$ outside the helium core \citep[e.g.,][]{Claeys11,Yoon17_Ib_IIb}. For these progenitors, a partial envelope ejection of $M_{\rm CSM}<0.1~M_\odot$ is not capable of powering a bright precursor of $\gtrsim 10^{41}$ erg s$^{-1}$ like the ones observed (except for the tightest separations; Figure \ref{fig:lc_properties}), whereas a complete envelope ejection would likely make the terminal SN identified as Type Ib (or IIn when CSM interaction happens) instead of Type IIb. Thus we expect bright precursors of SN IIb to be much less likely than SN IIn/II-P, which is consistent with the observations so far.}

\section{Conclusion}
\label{sec:conclusion}
We studied the emission from pre-SN outbursts from massive stars with a CO companion, aiming to reproduce the observed bright precursors of interacting SNe. We model the super-Eddington accretion of (a fraction of) the ejected CSM onto the CO, and the resulting energy injection into the CSM by the disk wind launched from the CO. We considered SN progenitors consisting of a RSG, a high-mass compact He star, and a low-mass extended He star (with parameters in Table \ref{tab:params}), each with a 10 $M_\odot$ BH or a 1.4 $M_\odot$ NS companion.

We constructed a one-zone light curve model including time-dependent energy injection, emission and acceleration of the CSM. We find that binary systems with separations of 3--30 times the stellar radii can generally reproduce both the luminosity and duration of the observed precursors, as well as the velocity of the CSM inferred from spectroscopy. 

For the RSG model motivated for precursors of hydrogen-rich (Type II) SNe, a BH companion leads to precursors with broad ranges of luminosity $10^{39}$--$10^{42}$ erg s$^{-1}$ and durations of months to years. We find that this can successfully explain the observed precursors, for a realistic range of CSM masses in Type II SNe corresponding to outburst energies only of the order of $10^{46}$--$10^{47}$ ergs. The case of NS companions can explain dimmer precursors of $\lesssim 10^{41}$ erg s$^{-1}$ which are observed in a fraction of Type IIn SNe \citep{Strotjohann21} but not the brightest ones approaching $\sim \! 10^{42}$ erg s$^{-1}$.

For the He star models motivated for Type Ibn SNe, the luminosity and duration of the precursors also agree with the two observed precursors of SN 2006jc and 2019uo. From the precursor luminosity, timescale, and the CSM velocity observed in the SN phase, we conclude that the progenitor of SN 2006jc is likely a compact (Wolf Rayet-like) He star with a BH companion, whereas that of SN 2019uo is likely an inflated He star with a NS companion. This may indicate a diverse progenitor channel for Type Ibn SNe, in addition to the claim that a fraction of these SNe may not be from massive stars \citep{Hosseinzadeh19}.

We find that the predicted event rates of our model roughly agree with the observed precursor rates, if enhanced mass-loss is relatively common months to years before core-collapse. However, we urge a more in-depth population synthesis study for rate predictions, as future observations will likely narrow down the uncertainty in the observed rate. Moreover we point out a possible link between the systems powering the precursors and CO binary systems observable by \textit{Gaia} and gravitational-wave detectors, which can also be investigated by such population synthesis studies.

In the era of the Rubin Observatory \citep{Ivezic19}, the number of SN precursors we can detect is expected to dramatically increase, and we may also be able to probe deeper into the dimmer end of the precursor population \citep[e.g.,][]{Tsuna23,Strotjohann24}. We expect that our framework would be helpful to characterize the origin and diversity of such precursors\footnote{\rev{We make our source code for calculating the light curves public in \cite{tsuna_binary_precursor}.}}. %(upon acceptance of this paper), accessible from: \url{https://github.com/DTsuna/binary_precursor}.}.

\begin{acknowledgments}
    We thank Xiaoshan Huang and Kareem El-Badry for stimulating discussions, and the referee for constructive comments that significantly improved this paper. D. T. thanks Morgan Macleod for discussions that led to inspiration of this model. D. T. is supported by the Sherman Fairchild Postdoctoral Fellowship at the California Institute of Technology. T. M. acknowledges supports from JSPS Overseas Research Fellowship and the Hakubi project at Kyoto University. S. C. W. is supported by the National Science Foundation Graduate Research Fellowship under Grant No. DGE-1745301. J. F. is grateful for support from the NSF through grant AST-2205974.
\end{acknowledgments}

\appendix
\section{Timestep Control for the Precursor Modeling}
\label{sec:dt_control}
When solving the governing equations for modelling the precursor described in Section \ref{sec:methods_with_recombination}, we carefully control the timestep of the light curve calculation so that the parameters in the governing equations do not suddenly evolve in a single step. Specifically, when $x_i=1$ the timestep is set by the dynamical and diffusion timescales as 
\begin{equation}
\Delta t_{x=1} = 10^{-2}{\rm min}\left(\frac{R_{\rm CSM}}{v_{\rm CSM}}, t_{\rm diff}\right).    
\end{equation}
When the wind is turned on with $x_i<1$, we impose another timestep limit using equation (\ref{eq:dxdt}) as 
\begin{eqnarray}
\Delta t_{x<1} 
={\rm min}\left[\Delta t_{x=1}, 10^{-2}t_{\rm diff}(5x_i x_{i,\bullet}) \left|\frac{L_{\rm inj}}{4\pi x_i^2 r^2\sigma T_I^4}-1\right|^{-1} \right].  
\end{eqnarray}
This makes $x_i$ evolve smoothly towards $x_{i,\bullet}$ (defined after equation \ref{eq:x_sat}), even when the wind is turned on when $x_i\ll 1$.

\section{Dependence on the Velocity Parameter $\xi$}
\label{sec:vary_xi}
Here we check the dependence of our results on $\xi (\gtrsim 1)$, the velocity of the ejected CSM normalized by the surface escape velocity of the progenitor, by conducting the same calculation as in the main text that assumed $\xi=2$ but for a reduced value of $\xi=1.4$. For a given time and fixed $M_{\rm CSM}$, the effect of reducing $\xi$ is to make the CSM less extended and denser. Thus we expect the energy injection by the disk wind to start later, and be shorter duration with higher power.

\begin{figure*}
     \centering
     \includegraphics[width=\linewidth]{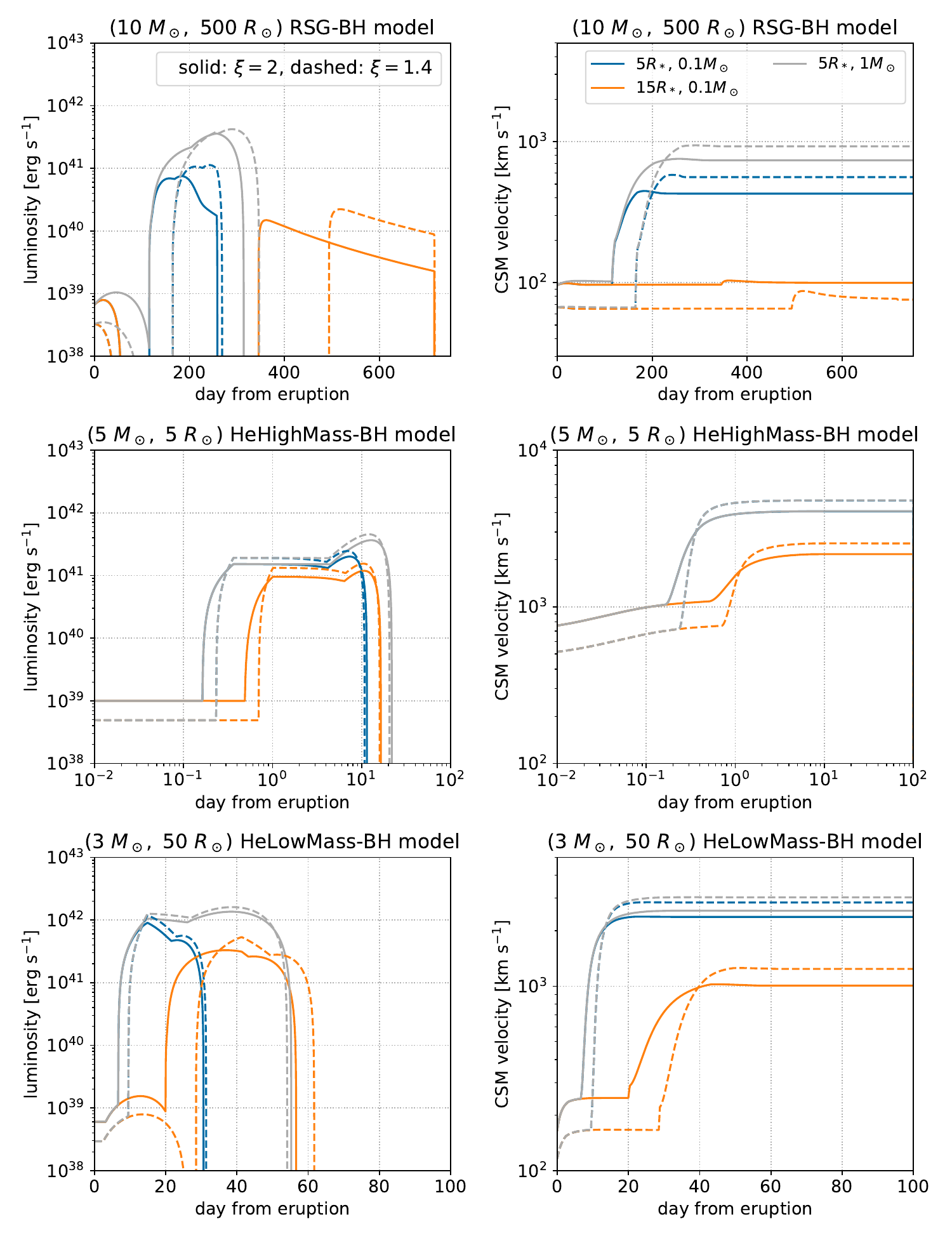}
     \caption{Dependence on the precursor light curves and final CSM velocities on the initial velocity parameter $\xi$. The solid lines are the case for $\xi=2$, and the dashed line are for $\xi=1.4$.}
     \label{fig:dependence_xi}
 \end{figure*}

Figure \ref{fig:dependence_xi} shows the dependence of the precursor light curves and final CSM velocities on $\xi$. We focused on the case for the BH companion, and adopted the same parameter sets as in Figure \ref{fig:few_cases_noepswind_BH}. As expected, for a fixed ($a_{\rm bin}, M_{\rm CSM}$) we see that for $\xi=1.4$ the precursor light curve rises at a later time than $\xi=2$, have a slightly brighter peak luminosity, and have a slightly faster final CSM velocity. However these quantities are very similar for the two cases of $\xi$, with differences typically within a few 10\%.

We note that the precursor duration can become quite different for extended stars with large $a_{\rm bin}$, as shown for the case of $a_{\rm bin}=15R_*$ for the top-left panel. This is because the precursor duration in this parameter region (typically 100s of days) is governed by the duration of the disk wind (equation \ref{eq:t_prec1}), rather than the diffusion timescale of the CSM. However since no precursors are observed for much longer than $100$ days, we expect this finding to not affect the discussion on the comparison of our model with the observed precursors in Section \ref{sec:result}.

\bibliography{references,references2}
\bibliographystyle{aasjournal}

\end{document}